%% file: main.tex
\documentclass[journal,twocolumn,a4paper]{IEEEtran}

\usepackage{latexsym}%
\usepackage{graphicx}%
\usepackage{multicol,multirow}%
\usepackage{amsmath,amssymb,amsfonts}%
\usepackage{mathrsfs}%
\usepackage{amsthm}%
\usepackage{rotating}%
\usepackage{ifpdf}%
\usepackage[T1]{fontenc}%
\usepackage{times}%
\usepackage{newtxmath}%
\usepackage{textcomp}%
\usepackage{xcolor}%
\usepackage{hyperref}%
\usepackage{lipsum}%
\usepackage{subcaption}%
\usepackage{orcidlink}%
\usepackage{algorithmic}
\usepackage{array}
\usepackage{textcomp}
\usepackage{stfloats}
\usepackage{hyperref}
\usepackage{url}
\usepackage{verbatim}
\usepackage{balance}
\usepackage{mathtools}
\usepackage[mathlines]{lineno}

\usepackage{nameref}

\usepackage{hyperref}
\hypersetup{
    unicode=false,          
    pdftoolbar=true,        
    pdfmenubar=true,        
    pdffitwindow=false,     
    pdfstartview={FitH},    
    pdftitle={},         
    pdfauthor={}, 
    pdfsubject={} 
    pdfcreator={}           
    pdfproducer={},         
    pdfkeywords={},         
    pdfnewwindow=true,      
    colorlinks=true,        
    linkcolor={blue},       
    citecolor=blue,         
    filecolor=blue,         
    urlcolor=blue           
}

\usepackage[backend=biber,
sorting=none,
doi=false,
isbn=false,
url=false,
maxcitenames=2,
mincitenames=1,
style=ieee,]{biblatex}
\begin{document}
\title{
Machine-learned Cloud Classes from Satellite data for Process-oriented Climate Model Evaluation}
\markboth{IEEE Transactions on Geoscience and Remote Sensing}
{Kaps
\MakeLowercase{\textit{(et al.)}:
Machine-learned cloud classes from satellite data for process-oriented climate model evaluation %
Process-oriented climate model evaluation using satellite data labelled with learned cloud classes }} 

\author{Arndt Kaps\orcidlink{0000-0002-5368-5950},
Axel Lauer\orcidlink{0000-0002-9270-1044},
Gustau Camps-Valls\orcidlink{0000-0003-1683-2138},~\IEEEmembership{Fellow,~IEEE,}
Pierre Gentine\orcidlink{0000-0002-0845-8345},
Luis Gómez-Chova\orcidlink{0000-0003-3924-1269},~\IEEEmembership{Senior Member,~IEEE,}
Veronika Eyring\orcidlink{0000-0002-6887-4885}
\thanks{A. Kaps (arndt.kaps@dlr.de), A. Lauer and V. Eyring with the Deutsches Zentrum für Luft- und Raumfahrt (DLR), Institut für Physik der Atmosphäre,  Oberpfaffenhofen, Germany}%
\thanks{G. Camps-Valls and L. Gómez-Chova with Image Processing Laboratory (IPL), University of Valencia, Valencia, Spain}%
\thanks{P. Gentine with Department of Earth and Environmental Engineering, Columbia University,  NY,  USA and Center for Learning the Earth with Artificial intelligence and Physics (LEAP),  Columbia University , NY,  USA}%
\thanks{V. Eyring with University of Bremen, Institute of Environmental Physics (IUP), Bremen,  Germany}
\thanks{
Funding for this study was provided by the European Research Council (ERC) Synergy Grant “Understanding and Modelling the Earth System with Machine Learning (USMILE)” under the Horizon 2020 research and innovation programme (Grant agreement No. 855187).
Gentine acknowledges support from National Science Foundation Science and Technology Center for Learning the Earth with Artificial intelligence and Physics (LEAP)
L.G.C. and G.C.V. work is partially funded by the Spanish Ministry of Science and Innovation (project PID2019-109026RB-I00).
The work was supported by the ESA Climate Change Initiative Climate Model User Group (ESA CCI CMUG).
This work used resources of the Deutsches Klimarechenzentrum (DKRZ) granted by its Scientific Steering Committee (WLA) under project ID bd1179.
The authors gratefully acknowledge the Leibniz Supercomputing Centre for funding this project by providing computing time on its Linux-Cluster.
We thank Valentina Zantedeschi, Fabrizio Falasca and their co-authors for kindly providing the \textsc{Cumulo} dataset and code.
Our own code related to the random forest and processing can be found at \texttt{\url{github.com/EyringMLClimateGroup/kaps22tgrs_ml_cloud_eval}} or via \texttt{DOI:10.5281/zenodo.7248773}}.
}
\maketitle

\begin{abstract}
Clouds play a key role in regulating climate change but are difficult to simulate within Earth system models (ESMs). Improving the representation of clouds is one of the key tasks towards more robust climate change projections. This study introduces a new machine-learning based framework relying on satellite observations to improve understanding of the representation of clouds and their relevant processes in climate models.
The proposed method is capable of assigning distributions of established cloud types to coarse data. It facilitates a more objective evaluation of clouds in ESMs and improves the consistency of cloud process analysis. The method is built on satellite data from the MODIS instrument labelled by deep neural networks with cloud types defined by the World Meteorological Organization (WMO), using cloud type labels from CloudSat as ground truth. The method is applicable to datasets with information about physical cloud variables comparable to MODIS satellite data and at sufficiently high temporal resolution. We apply the method to alternative satellite data from the Cloud\_cci project (ESA Climate Change Initiative), coarse-grained to typical resolutions of climate models. The resulting cloud type distributions are physically consistent and the horizontal resolutions typical of ESMs are sufficient to apply our method. 
We recommend outputting crucial variables required by our method for future ESM data evaluation. This will enable the use of labelled satellite data for a more systematic evaluation of clouds in climate models.
\end{abstract}%

\IEEEpeerreviewmaketitle
\section{Introduction\label{sec:Intro}}

\IEEEPARstart{E}{arth} system models (ESMs, also referred to as climate models) are important tools not only to improve our understanding of present-day climate but also to project climate change under different plausible future scenarios. 
 
The simulation of clouds and their interactions with the climate system, however, remain a major challenge for ESMs \cite{Vignesh2020}.
The representation of clouds in these models has been identified as one of the primary sources of inter-model spread \cite{Dufresne2008,Zelinka20}. An improved representation of cloud processes in ESMs is therefore an essential component in addressing these issues \cite{Bony2015,Schneider2017,Williams2009}.

Observations frequently used to assess model performance are obtained from long-term satellite products providing near-global coverage, which have proven to be well suited for the evaluation of climate models \cite[e.g.][]{Lauer2017,gmd-12-2875-2019}. This conventional approach is, however, constrained in part due to limitations and uncertainties of observational products themselves \cite{jakob2003improved}, such as biases or varying spatial and temporal coverage. 

We propose a new approach to ESM evaluation, designed to facilitate process-oriented evaluation of clouds in climate models and to address some of the apparent limitations of using conventional observational data. We use \emph{a priori} knowledge about the characteristics of different cloud classes based on the cloud type classification of the World Meteorological Organization (WMO).  By exploiting this \emph{a priori} knowledge, cloud processes can be highlighted in further evaluation. Our approach extends the recent development of machine learning based cloud classification methods for satellite data \cite{Rasp2019,Zantedeschi,Denby2020,Kurihana2021,Zhang2019,marais2020} to climate models. Machine learning-based cloud classification is not a new idea \cite[e.g.][]{Lee1990}, but has only recently become feasible for large-scale applications due to the increase in available computing power and the different available methods have distinct properties. An important distinction between classification methods is whether they are supervised or unsupervised. The former relies on previously assigned classes and the latter aims at automatically finding distinct new classes. Supervised classification relies on the assumption that the assigned classes fit the purpose, whereas the user has limited control over the makeup of the classes in unsupervised methods. Therefore, supervised methods allow for interpreting the final results without additional analysis steps but require a set of labelled data \cite{Zantedeschi,Zhang2019}. If the goal is to find classes that are as distinct as possible, or if no previously labelled data are available, unsupervised methods are preferable \cite{Kurihana2021,Denby2020}.

To our knowledge, no high-resolution ($\mathcal{O}(1\;\rm{km})$) cloud-class-labelled satellite data have been used for analysis and evaluation of ESMs, so far. labelled datasets allow for a more detailed and more direct interpretation of cloud classes in the respective satellite data in contrast to comparatively coarse classifications as used for example in the D-Series of the International Cloud Climatology Project (ISCCP, \cite{Rossow1999}). Previous studies have used clustering for satellite products and the output of satellite simulators from models for unsupervised classification of clouds \cite{Williams2009,Jakob2003,Gordon2005}. In these studies, morphological cloud regimes are then assigned to the identified clusters, according to the average physical properties of each cluster. Such a classification offers valuable insights into how individual models represent clouds in a more specific way than simple climatologies of physical variables would. However, in addition to being built on the rather low resolution of $(280\;\rm{km})^2$ of the ISCCP-D1 \cite{Rossow1999} product, uncertainties and artifacts introduced by the satellite simulators can affect the results \cite{Pincus2012,Swales2018}. In a recent study, a convolutional neural network was used on $(4000\;\rm{km})^2$ grid cells to assign the amount of each of four cloud classes per cell \cite{Kuma2022}. In \cite{Kuma2022}, the classes were derived from WMO classes detected from surface observations, and the method is applicable to climate model output.  Other studies have classified satellite data by cloud regime to investigate specific cloud properties such as radiative effects or precipitation \cite{Oreopoulos2016,Tan2015}. Recently, clustering methods for cloud regimes have been applied to current-generation climate models using the $1^{\circ}\times 1^{\circ}$ resolution ISCCP-H product \cite{Young2018} for training, which has a much higher resolution than ISCCP-D1 \cite{Tselioudis2021}. Most of the clusters found this way are labelled with cloud regimes named after cloud types defined by the WMO.

Our method aims at establishing this connection between observations and models without the requirement to assign cloud classes \emph{a posteriori}. We instead compute the relative amount of WMO cloud classes in coarse grid cells as explained in Section \ref{sec:meth}. Statistical analysis can be conducted on these distributions in the same manner as for the traditionally used physical variables but in the phase space of cloud classes.

The proof-of-concept is outlined as follows. In section~\ref{sec:meth}, we describe the satellite products used and introduce the two machine learning methods applied. In section~\ref{sec:res}, we use our results to establish (1) that a neural network can be used to accurately assign cloud class labels to satellite data, (2) that this labelled satellite data provides a sufficient basis to train a regression model relating physical variable retrievals from satellites to cloud class distributions in coarse grid cells and (3) that the application to a coarse-grained version of the alternative ESA Cloud\_cci satellite product \cite{Stengel2017} is possible, showing the potential of the framework. Our findings are summarized in Section \ref{sec:summ}. A discussion and outlook are presented in section~\ref{sec:Dis}.

\section{Methods\label{sec:meth}}
\subsection{Overview}

\begin{figure}
    \centering
    \includegraphics[width=\linewidth]{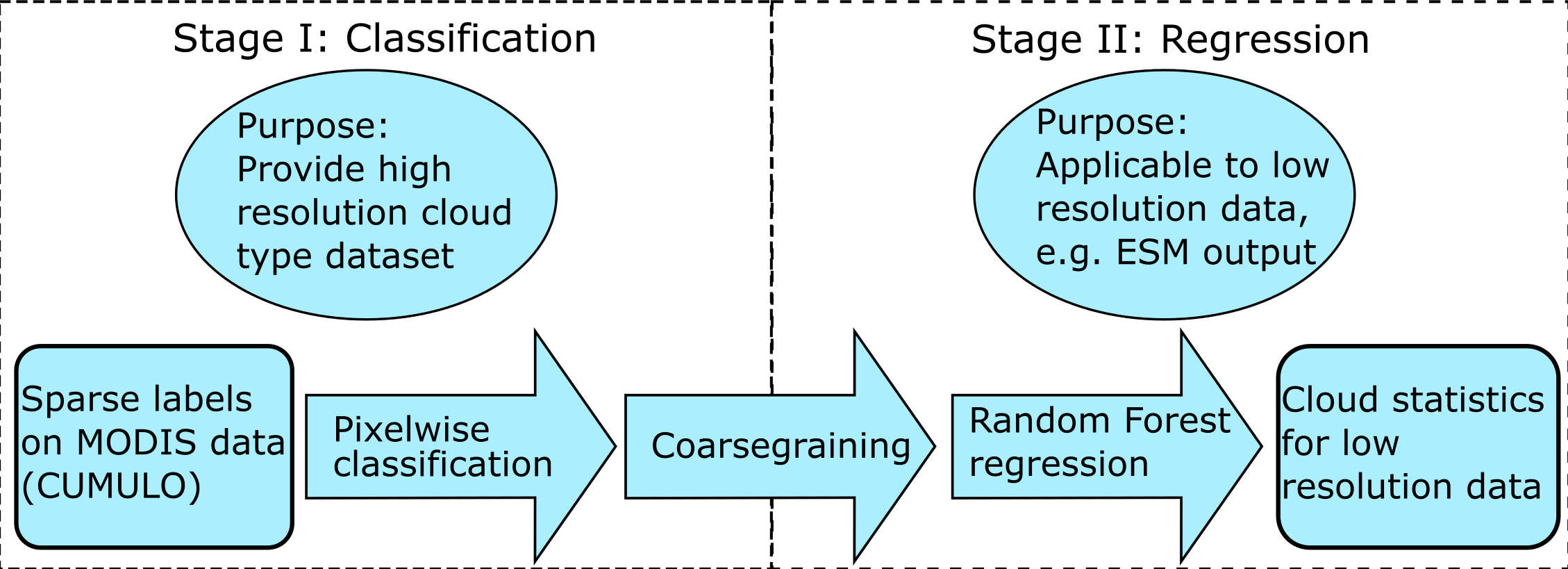}
    \caption{Two stages of machine-learning - a classifier and a regression model - are required to obtain cloud type predictions on datasets with low horizontal resolution. }
    \label{fig:concept}
\end{figure}

\begin{figure*}
    \centering
    \includegraphics[width=0.8\linewidth]{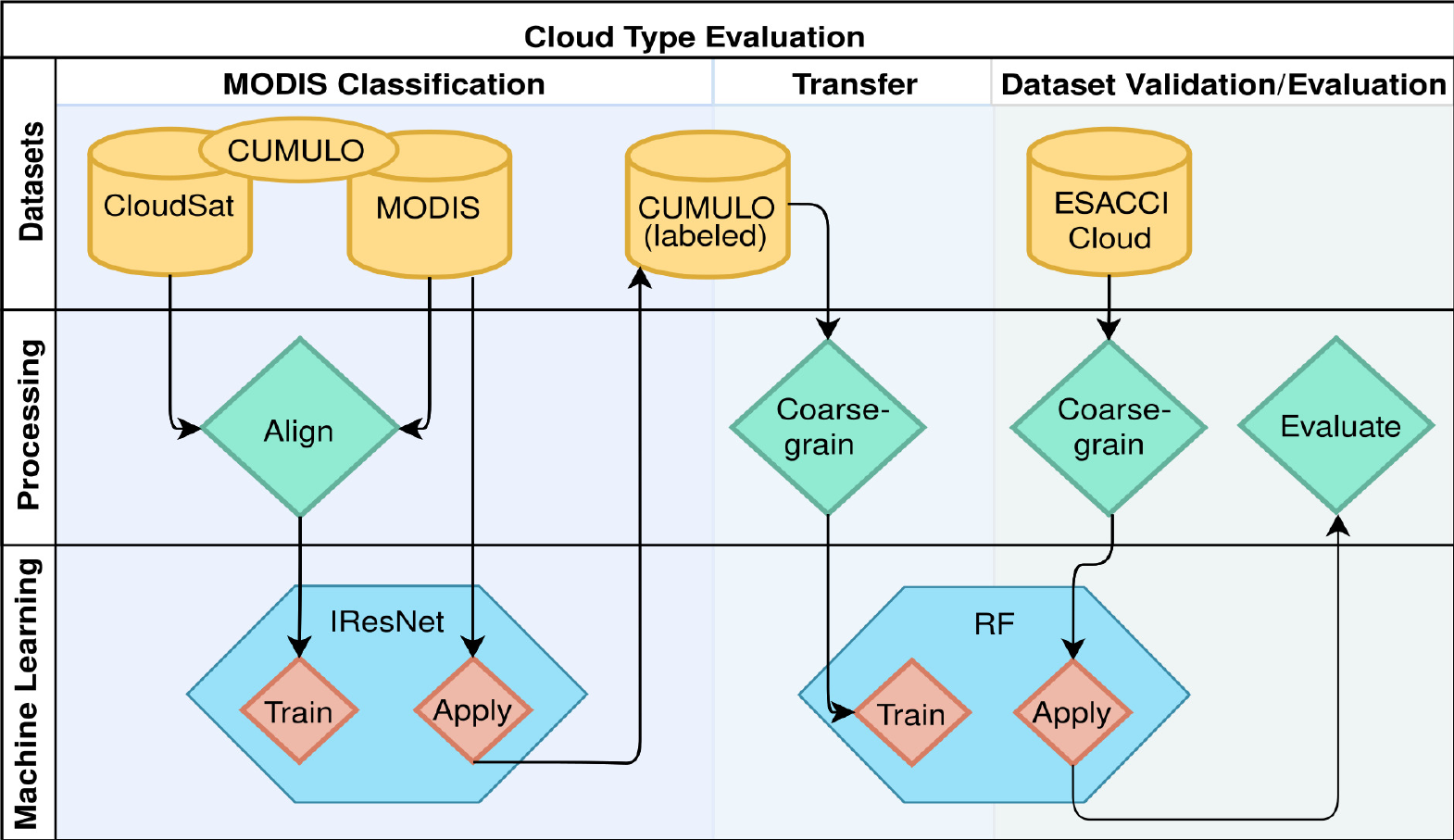}
    \caption{Workflow schematic: (1) The IResNet is trained on the \textsc{Cumulo} dataset  and then applied on the unlabelled full-swath MODIS yielding the fully labelled \textsc{Cumulo} dataset. (2) A random forest (RF) regression model is trained on a coarse-grained version of this data to provide cloud class distributions. (3) The RF is applied to unseen data, allowing validation of the methods performance or evaluation of the target dataset.}
    \label{fig:schematic}
\end{figure*}

Our goal is to evaluate clouds in climate models using observational data labelled with cloud types. For this, we need to
\begin{enumerate}
    \item obtain or create a cloud-labelled dataset 
    \item enable a comparison to climate model output
\end{enumerate} 
As a starting point, we use the partially labelled \textsc{Cumulo} dataset \cite{Zantedeschi}, which is then fully labelled by using a neural network classifier. As this dataset has much higher horizontal resolution than typical climate model output this classifier can not be used for climate models directly. Instead, we train a regression model to predict the relative amounts of each cloud type that are present in larger areas (grid cells) of the fully labelled dataset. This concept is outlined in Fig. \ref{fig:concept}. \\
In Section \ref{sec:meth_class} (Stage I in Fig. \ref{fig:concept}), we outline how the \textsc{Cumulo} dataset was created using data from the Moderate Resolution Imaging Spectroradiometer (MODIS) and cloud type labels from CloudSat. Since classifying low resolution data with individual cloud types is not appropriate, a regression model is trained on a coarse-grained version of this dataset (Section \ref{sec:meth_reg} (Stage II in \ref{fig:concept}). 
Section \ref{sec:meth_eval} explains the steps applied to validate the regression model's performance (see Fig. \ref{fig:validation}) on coarse-grained data from ESA Cloud\_cci, which are independent of the training data. In this step, we obtain cloud class distributions, i.e. the percentages of each cloud type in each coarse grid cell of the dataset. Model datasets can be evaluated by relating these distributions to the underlying processes driving formation and evolution of the specific cloud type. 

The workflow of this framework is shown in Fig. \ref{fig:schematic}, which illustrates how the different datasets and the two machine-learning models contribute to providing cloud type distributions for low resolution data.

\subsection{Satellite data products\label{sec:data}}
\begin{table*}[]
    \centering
    \begin{tabular}{c|c|c|c|c|c}
         Name & Product / Version& Purpose & Resolution & Revisit period & Reference  \\
         \hline 
         CloudSat   &  2B-CLDCLASS-LIDAR / P1-R05     &  cloud type     &  $(1.4\times 1.8)\;\rm{km}^2$      &   16 days    & \cite{PCICD2019},\cite{Sassen2009} \\
         \hline 
         MODIS          & MYD06 / 6.1      &   physical variables   & 1\;km     & 1-2 Days      &  \cite{Platnick2003}\\
         \hline 
         ESA CCI        & ESA Cloud\_cci  L3U  / AVHRR-PMv3   &  physical variables for validation     &  0.05°     &  ~$\frac{1}{2}$ day     & \cite{Stengel2020}
    \end{tabular}
    \caption{The three satellite datasets used at different stages in our framework.}
    \label{tab:sats}
\end{table*}

This work is based on data from three satellite products summarized in Table \ref{tab:sats}. 
\subsubsection{CloudSat}
The cloud classes used in the \textsc{Cumulo} dataset are obtained from  CloudSat. The product combines radar measurements from CloudSat and lidar data from the Cloud-Aerosol Lidar with Orthogonal Polarization (CALIOP) of the Cloud-Aerosol Lidar and Infrared Pathfinder Satellite Observation (CALIPSO) satellite \cite{Winker2006}. These are used as the basis for a mixed threshold-based/fuzzy logic cloud classifier. 
The use of active sensors provides data that uniquely enable the labelling of multiple clouds along the vertical. A limitation of this dataset is its sparse spatial and temporal coverage. Also, the small footprint size of the satellite instruments used makes distinction between stratocumulus and stratus clouds difficult. 

\subsubsection{MODIS}
The inputs (called features in the following) to the machine-learning algorithms used here are also included in \textsc{Cumulo} and obtained from the Cloud Product of the MODIS instrument, which operates aboard the Terra and Aqua satellites. These are sun-synchronous polar-orbiting satellites like CloudSat and CALIPSO, which together with Aqua were part of NASAs afternoon constellation (A-Train) from 2006 to 2018, providing near simultaneous measurements. The MODIS Cloud Product data include nine pixel-level retrievals of physical variables (Table~\ref{tab:vars}) as well as 13 radiance channels. These variables are provided for images of 1354$\times$2030  $1\;\rm{km}^{2}$ pixels each covering five minutes. 

Known limitations include large uncertainties in the detected cloud phase in high elevation regions, including Greenland and Antarctica. We use the MODIS data for the year 2008 provided with the\textsc{Cumulo} dataset.

\subsubsection{ESA Cloud\_cci}
For validation of the method, we use the ESA Cloud\_cci (ESA CCI) dataset. It is a long-term cloud product obtained from different observational sources. We use the dataset based on data from the Advanced Very High Resolution Radiometer (AVHRR) aboard the polar orbiting afternoon satellites from the National Oceanic and Atmospheric Administration (NOAA). Among others, the retrieved variables include a cloud mask and cloud physical variables that are also available from the MODIS Cloud Product (Table~\ref{tab:vars}). The retrievals included in the ESA CCI data are obtained using the Community Cloud retrieval for Climate (CC4CL) algorithms \cite{Sus2018}. CC4CL shows little dependence on the observational instrument or the surface properties of the measurement region, but still suffers from some particular issues for retrievals using passive sensor data. This includes difficulties in assessing the thermodynamic phase of mixed-phase clouds as well as the detection of thin cirrus clouds. We use the daily L3U dataset for June 2009 up to and including December 2011, with July 2010 being removed due to faulty temperature retrievals.

\subsection{Pixel-wise classification\label{sec:meth_class}}
\begin{figure}
    \centering
    \includegraphics[width=\linewidth]{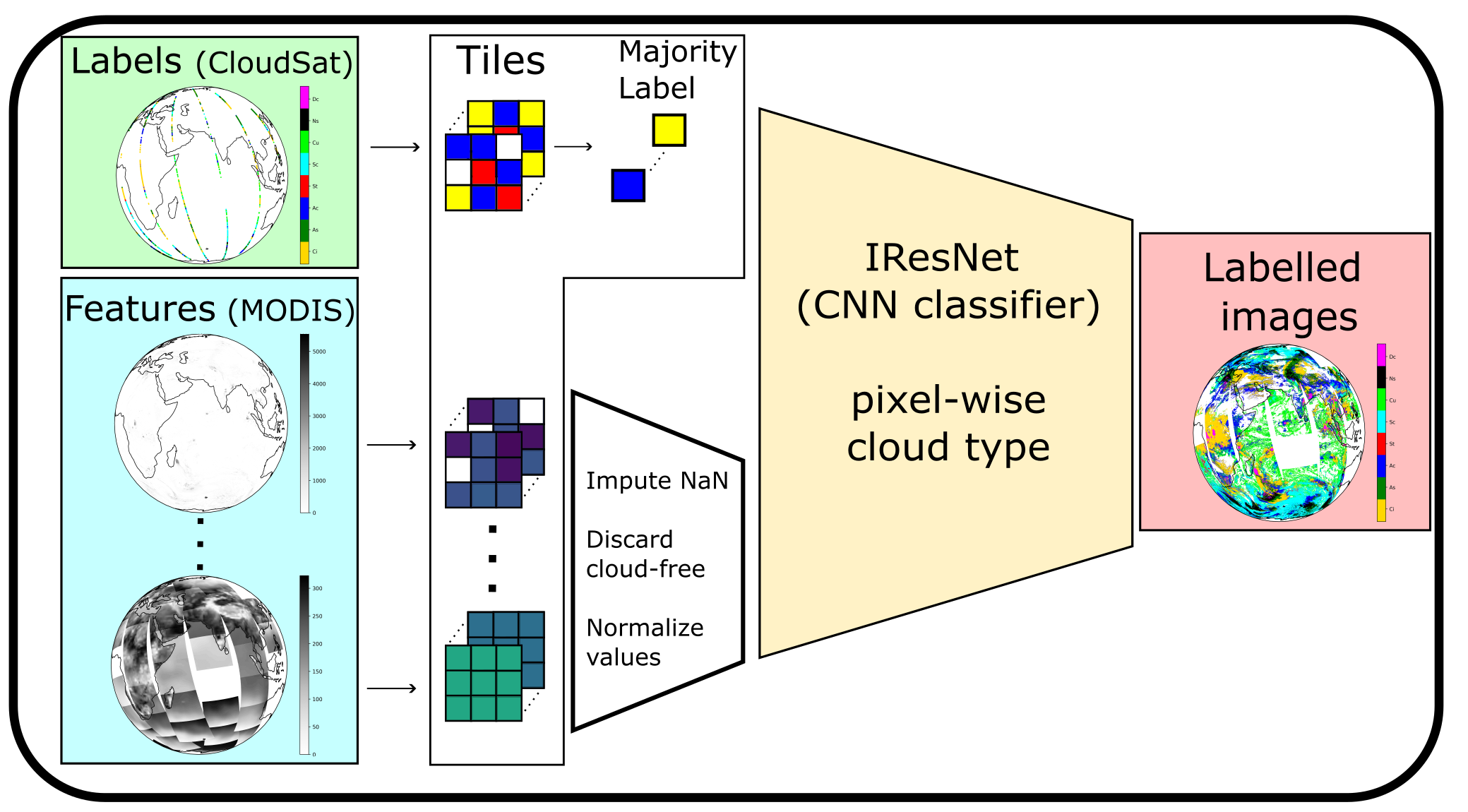}
    \caption{Schematic of the pixel-wise classifier, which is a convolutional neural network trained on features from MODIS and one of eight cloud type labels from CloudSat per pixel.}
    \label{fig:iresnet}
\end{figure}

We create a fully labelled cloud type dataset by applying a pixel-wise classifier network to the sparsely labelled \textsc{Cumulo} datset \cite{Zantedeschi}. The results is a high-resolution, high-coverage, cloud-labelled dataset (see Fig. \ref{fig:iresnet}). The classification scheme applied here is largely based classification algorithm used in \cite{Zantedeschi}. The \textsc{Cumulo} dataset provides one of eight cloud type labels (see Table~\ref{tab:ctypes}) per cloudy pixel along a narrow, vertically resolved path. To be compatible with the 2-dim MODIS data, each vertical column is represented by the class which occurs most often in that column. The data from the two satellites are aligned such that, where available, the label track from CloudSat is superimposed on the corresponding MODIS data. This excludes nighttime measurements, as some MODIS retrievals are not available at night. Since the CloudSat swath is quite narrow, most of the pixels in the resulting \textsc{Cumulo} dataset are not assigned a label, which is why a neural network is trained using the labelled part of the dataset to predict cloud class labels for the unlabelled pixels. \\
\begin{table}
    \caption{Physical variables available from the MODIS Cloud Product, usable as features for the machine learning models.}
    \label{tab:vars}
    \centering
    \begin{tabular}{c|c}
           Abbv. & Physical variable  \\
           \hline
         \textit{cwp} & Cloud water path\\
         \textit{cod} & Cloud optical thickness\\
         \textit{ptop}& Cloud top pressure\\
         \textit{htop}& Cloud top height\\
         \textit{ttop}& Cloud top temperature\\
         \textit{tsurf}& Surface temperature\\
         \textit{cer}& Effective cloud particle radius\\
         \textit{ceff}& Effective Emissivity\\
         \textit{phase}  & Cloud thermodynamical phase
    \end{tabular}
\end{table}
\begin{table}
    \caption{Cloud types from the \textsc{Cumulo} dataset.}
    \centering
        \begin{tabular}{c|c}
            Abbv. & Cloud type \\
            \hline
             Ci & Cirrus/Cirrostratus   \\
             As & Altostratus \\
             Ac & Altocumulus \\
             St & Stratus \\
             Sc & Stratocumulus  \\
             Cu & Cumulus  \\
             Ns & Nimbostratus  \\
             Dc & Deep Convection  \\
        \end{tabular}
        \label{tab:ctypes}%
\end{table}\\
The neural network used in \cite{Zantedeschi} and here to classify clouds in the \textsc{Cumulo} dataset is a semisupervised convolutional network based on the invertible residual network (IResNet) \cite{Behrmann2019}. Residual networks \cite{He_2016_CVPR} have become the baseline for many image-related tasks and the IResNet additionally allows for semisupervised training. The training is termed semisupervised as both labelled and unlabelled samples are fed to the network. The model learns to minimize the cross-entropy for the labelled parts (Eq. \ref{eq:labelled}) as well as the negative log-likelihood (Eq. \ref{eq:unlabelled}) of the latent representation $z$ of all (labelled and unlabelled) samples.\\
\begin{linenomath*}
\begin{align}
    \mathcal{L}&=-\sum_{ \mathclap{\substack{z_k = F(x_k),\\ x_k \in \mathcal{X}}}}  \log(p(z_k)) + \mathrm{Tr}(\mathbf{J}_F)\label{eq:unlabelled},\\
    \mathcal{L}_l &= \sum_{x_k,y_k \in \mathcal{X}_l} log(x_k)y_k .\label{eq:labelled}
\end{align}
\end{linenomath*}
Here, $z_k=F(x_k)$ is the latent output of the IResNet $F$ without its classifier head, and $\mathbf{J}_F$ is its Jacobian, with $Tr$ denoting the trace operation. $\mathcal{X}$ contains all samples $x$, $\mathcal{X}_l$ contains only the samples with labels $y$.\\ 
The IResNet is applied to tiles consisting of $3\times 3$ MODIS pixels to determine the cloud type of the central pixel. The training target is the cloud class that occurs most often in the tile or a random choice of classes that occur equally often in the tile. Note that this way, the class label is predicted such that it is representative of the whole tile, even though the label is only assigned to the central pixel. This is a design choice that possibly introduces a bias towards more frequent cloud classes but increases the number of usable tiles both for training and prediction by allowing for overlapping tiles. Tiles that contain less than six cloudy pixels according to the MODIS cloud mask are discarded. Therefore, the neural network is agnostic to such cases including clear sky situations. By applying the trained model, pixels in the \textsc{Cumulo} data that are yet unlabelled are assigned class labels, resulting in a set of fully labelled satellite data.

The \textsc{Cumulo} data contain MODIS radiance channels as well as retrieved physical cloud properties (Table~\ref{tab:ctypes}). With potential application to climate models in mind, we decided to train the IResNet using the physical variables as features, these being more readily available from climate models than the radiances at the particular MODIS spectral channels. We found that the classes predicted by the model trained on the physical variable features were slightly more physically consistent. For example, we found that a number of high and thin clouds were given the Cumulus (Cu) label when using the radiances only, but this did not happen when using the physical cloud properties. However, the performance difference was marginal such that the classification step could be trained on either set of features. To be able to perform the training on physical variables, pixels containing missing values e.g. from failed MODIS retrievals are imputed, using the mean value for each $3\times 3$ pixel tile. As the tiles are small, this is not expected to skew the values in the individual tiles significantly as neighboring pixels are expected to have similar properties.

The IResNet is trained on all available \textsc{Cumulo} granules for the year 2008 ($\sim 48000$ multivariate images of $1354\times 2030$ pixels) with standardized features. Instead of using a train/test split, we used 5-fold cross-validation on the same data to assess generalization to unseen data. The model used for final predictions is then trained on the complete year. Due to the high temporal resolution of the data, there should be enough variance in the features for the training data to be representative of longer periods typical for climate models. This will compensate for the fact that only one year of training data is used.

\subsection{Regression on low resolution data\label{sec:meth_reg}}
\begin{figure}
    \centering
    \includegraphics[width=\linewidth]{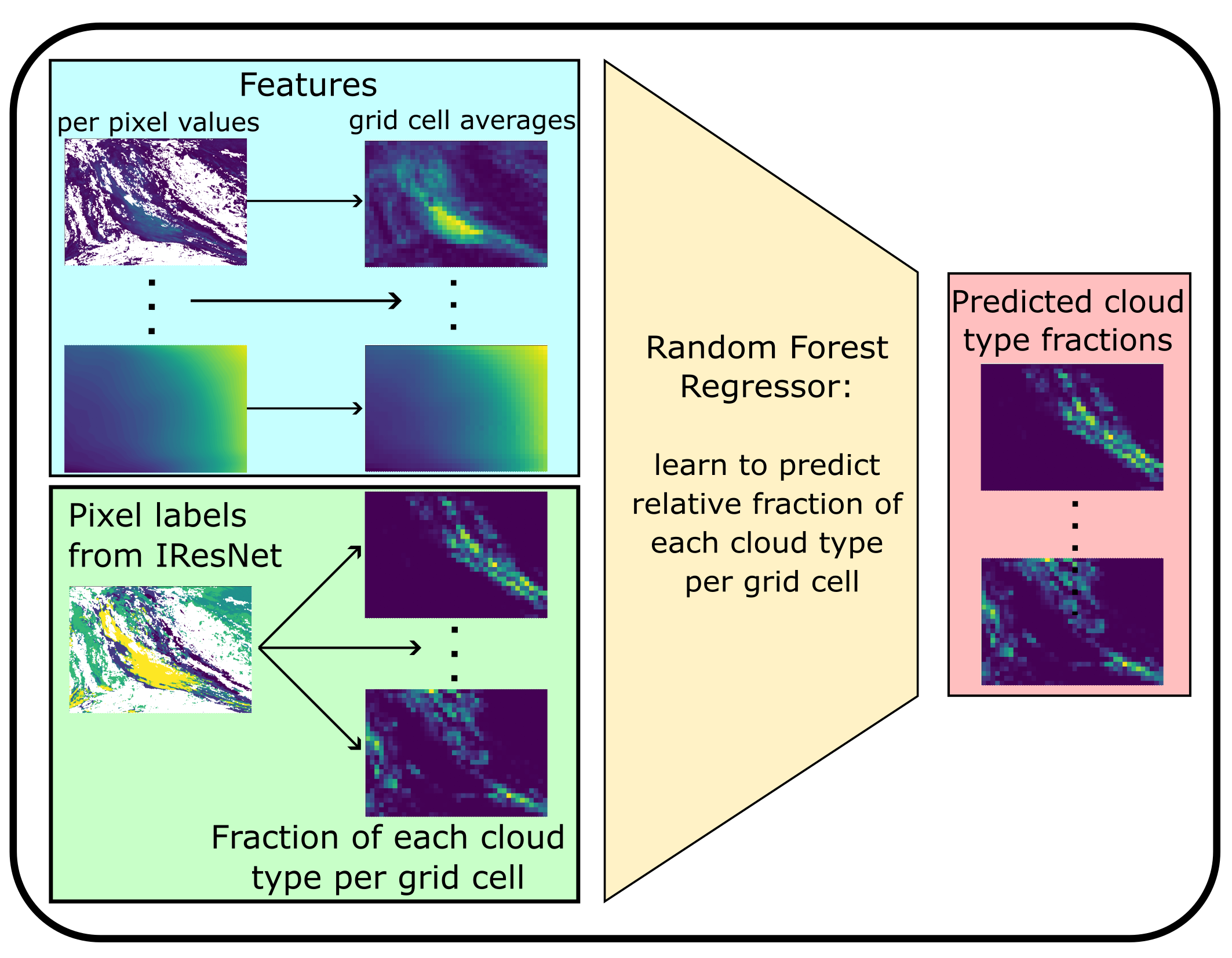}
    \caption{Cloud type predictions for data with low horizontal resolution are obtained by coarse-graining high resolution predictions as a basis to train a regression model predicting relative amounts of each cloud type for each coarse resolution grid cell.}
    \label{fig:RF}
\end{figure}
The second stage of machine learning (see Fig. \ref{fig:RF}) is designed to transfer the information contained in the (labelled) high-resolution satellite data to datasets of lower temporal and spatial resolution, like typical ESM output. 

For this, the labelled satellite dataset obtained from the pixel-wise classification is coarse-grained. All variables that are provided in both the \textsc{Cumulo} and the target dataset can be used as features. 

The labelled data are provided on an evenly spaced metric grid, but many climate models are provided on evenly spaced angular grids. The area covered by individual pixels will not match between these two grids and scale differently depending on their geographic location. For simplicity and for the purpose of a proof-of-concept, we determine the grid cell size that on average is most representative of the target grid and use the averages of each variable over these grid cells as features for our model. We assume that the remaining differences between the grids are mitigated by averaging. The output is the relative cloud class occurrence in the grid cell, i.e. the fractional amount of each of the eight cloud classes plus an additional ``undetermined'' class. The ``undetermined'' class contains all pixels for which the prediction of a label was not possible due to failed MODIS retrievals, which often indicate clear sky. Missing values for pixels with no cloud are processed accordingly when computing the grid-cell averages (see section~\ref{sec:input}), such that cloud class fractions are predicted consistently for all properties including those that are not defined for clear sky (e.g. \textit{ptop}). Grid cells containing only ``undetermined'' pixels are discarded. Thus, we obtain a multivariate regression problem with a nine-dimensional output space, containing the eight classes plus ``undetermined'' pixels, and up to eleven features (i.e. the number of suitable physical variables provided by the \textsc{Cumulo} data, see section~\ref{sec:input}).

For our model, we choose the random forest (RF) \cite{Breiman2001} regression method for reasons of simplicity, computational efficiency, as well as its inherent normalization of the predicted fractions. After training the RF on the coarse-grained classified images, it can be directly applied to the target data, i.e. ESM output, providing cloud class fraction predictions for each grid cell. In order to investigate the sensitivity to the resolution and choice of features used, we trained multiple RFs for different respective choices. The individual training samples are weighted with weights $w_i$ given by the $L^1$-norm $w_i = ||\mathbf{y}_i- \overline{\mathbf{y}}||_1$,
where $\mathbf{y}_i$ denotes the cloud class fractions for the $i$-th training sample and $\overline{\mathbf{y}}$ the average over all samples used in training. The weighting ensures that samples close to ``the average sample'' are given less weight in training, to reduce the effect of any bias in the data. We have about 48,000 labelled \textsc{Cumulo} granules (multivariate images of $1354\times 2030$ pixels) available. In order to limit the amount of memory required, the RFs are trained on roughly $50\cdot 10^6$ random samples drawn from a training split of $10000$ labelled data images. The amount of samples varies because grid cells containing only ``undetermined'' pixels are excluded. The models are then evaluated on a test split containing 8422 images. The hyperparameters of the RF models are chosen such that the depth of the individual regression trees is seventeen or less. We apply a bagging subsampling fraction of $0.7$ and a minimum leaf size of $2$, with $400$ individual trees per forest. These hyperparameters showed an optimal trade-off between the model's performance and size.

\subsection{Application to ESA CCI data \label{sec:meth_eval}}

\begin{figure}
    \centering
    \includegraphics[width=\linewidth]{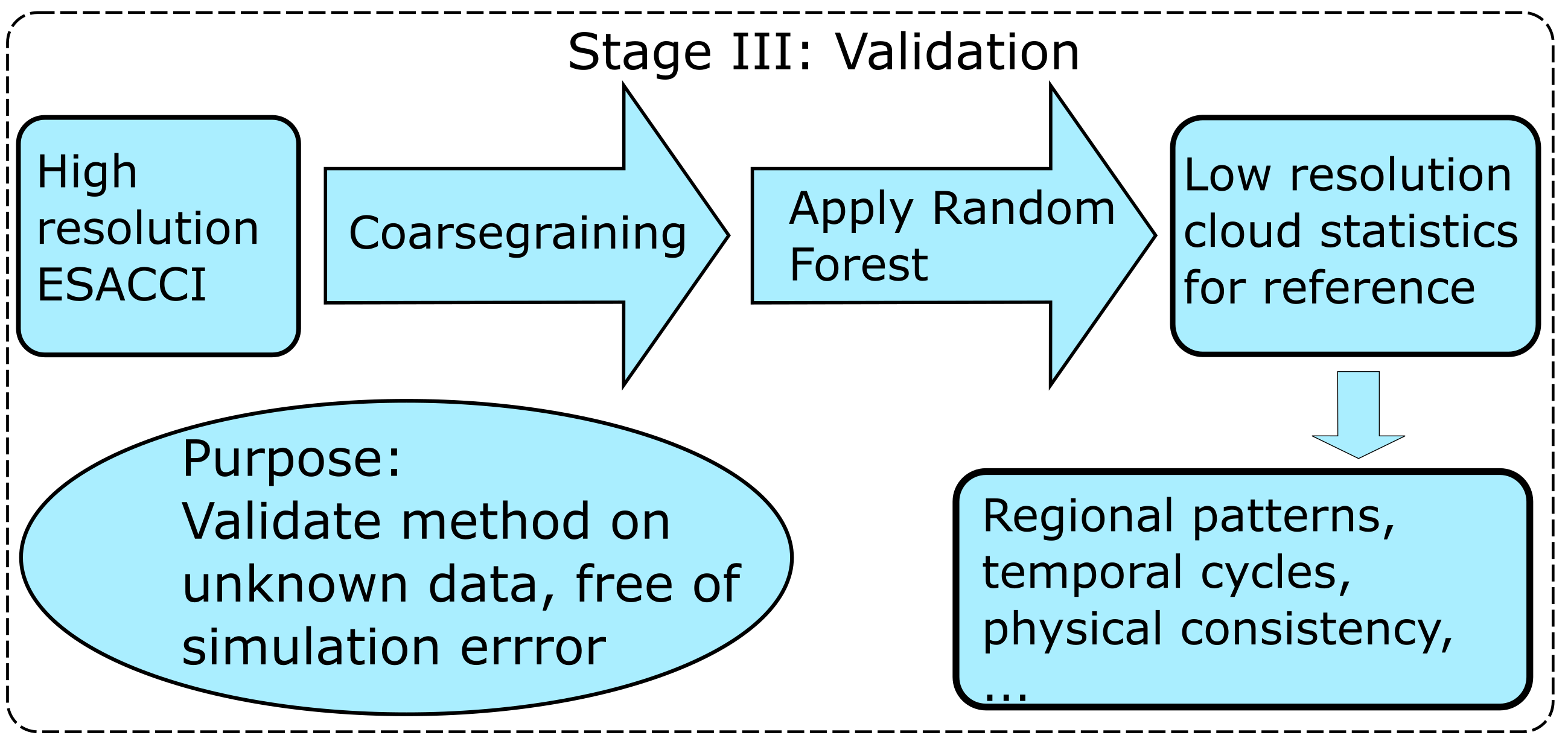}
    \caption{The method is validated by applying the trained regression model to data the model has not seen before. The predictions are then analysed for physical consistency.}
    \label{fig:validation}
\end{figure}

As a proof-of-concept, we apply our method to an independent satellite dataset, the ESA CCI data (see Section~\ref{sec:data}). Application to the output of current ESMs (e.g. those contributing to CMIP6) is not yet possible due to too coarse horizontal and/or temporal resolution of the available output and/or key variables (e.g. \textit{cer}, \textit{cod}), required for sufficient performance, being unavailable. As illustrated in Fig. \ref{fig:validation}, we use this validation stage to show that the method generalizes to coarse data obtained from different sources and it is thus expected to also be applicable to other datasets such as suitable ESM output.

The ESA CCI data provide a similar representation of the observed cloud state and contain similar physical cloud variables as the MODIS product and are therefore comparable to the \textsc{Cumulo} data. In contrast to using ESM output as target data, this allows for a more direct assessment of uncertainties and limitations of this approach as possible model deficiencies do not play a role. We provide an analysis of predictions using different spatial and temporal resolutions in the supplementary material.
For each validation experiment using the coarse-grained ESA CCI data, we randomly sample 20\% of the available grid cells.

\subsection{Features and preprocessing\label{sec:input}}
The RF regression model is trained on the features also available in the target dataset. For the ESA CCI data, these include cloud top temperature (\textit{ttop}), cloud top height (\textit{htop}), cloud top pressure (\textit{ptop}), surface temperature (\textit{tsurf}), cloud optical thickness (\textit{cod}), cloud water path (\textit{cwp}) and the effective radius of cloud particles (\textit{cer}). For the cloud water path and the effective radius, the MODIS cloud product as well as the ESA CCI daily product do not distinguish between ice and liquid water. Instead, an additional flag is available to distinguish between liquid and ice cloud tops, which we use to separate \textit{cwp} into liquid water path (\textit{lwp}) and ice water path (\textit{iwp}) as well as \textit{cer} into the liquid and ice particle radii (\textit{cerl/ceri}). This procedure is an approximation and only justified under the assumption that the phase flag provided by the satellite data is representative of the whole cloud column. 

The grid box averages for the cloud liquid/ice water path, the radii, \textit{cod} are computed over all pixels in each cell, i.e. replacing missing values with zero. This is useful as these values approach zero with decreasing cloud amount. In contrast, \textit{ptop, htop, ttop} are only averaged over cloudy pixels (``in-cloud values'').
The features used for the RF model should ideally complement each other. Information on the cloud thickness is implicitly contained in both \textit{cwp} and \textit{cod}. As the features \textit{ptop, htop} and \textit{ttop} effectively contain the same physical information, only one of them is used. Here, we use \textit{ptop} as feature. In addition to the cloud variables, we also use surface temperature \textit{tsurf} as it is readily available in many datasets. As a default, we therefore select \textit{cwp, lwp, iwp, cerl, ceri, cod, ptop, tsurf}, which we call the \emph{optimal set} of features in the following.

The ESA CCI dataset also provides a pixel-wise uncertainty estimate for each variable, which we use to exclude pixels for which the uncertainty is larger than twice the actual value. We note that \textit{cwp} in the ESA CCI data can take very large values. Such outliers with \textit{lwp}$>2000\;\rm{g/m^2}$ or \textit{iwp}$>6000\;\rm{g/m^2}$ are excluded from the prediction. The arithmetic means of the latitude and longitude coordinates of all pixels in a grid cell are used as representative geographical locations.

\section{Results\label{sec:res}}
\subsection{Predicted cloud classes at pixel level\label{sec:res_class}}
To assess the performance of the IResNet we use accuracy and F1-Score (see Table~\ref{tab:histo}). A qualitative analysis of the physical properties of the predicted cloud classes is used to evaluate the consistency of the results. This is important because the physical properties of the classes predicted by the IResNet model need to be consistent with those from the WMO definitions. 

The labels extracted from CloudSat that are available in \textsc{Cumulo} display a strong class imbalance (Table~\ref{tab:histo}), which we also find in the predicted classes. Most classes occur with a similar frequency in the source data and predictions, with deviations being small enough to be attributable to real differences in the data. We would like to highlight two key properties of the class distributions: (1) there are very few strati (St) and deep convective (Dc) clouds in both the source and the prediction. (2) Cumulus (Cu) and cirrus (Ci) clouds are strongly underestimated in the predictions compared to the source data. For example, Cu has the smallest amount of all predicted cloud classes while this class is more common than St and Dc in the source data. Further assessment of the representation of these four classes (St, Dc, Cu, Ci) is therefore of high importance. The mean accuracy of the classification in the validation splits of the cross-validation is larger than $0.8$ for all classes but Sc, which suggests considerable skill in the classification method. However, for a multi-class problem with a large class imbalance such as the one we have here, the accuracy alone is not a suitable measure to fully assess the performance of the method. We therefore additionally look at the F1-Score, which is sensitive to the class imbalance and can help identify individual class biases in the predictions. The F1-Score is at least $0.4$ for all classes except for St, and especially high for Sc and Cu with values larger than $0.6$. When considering the accuracy scores for all classes, this suggests a good skill in representing the class imbalance.  An exception is the St class with an F1-Score of $0.21$. As noted in section~\ref{sec:data}, the CloudSat algorithm has trouble distinguishing between St and Sc, which is why this is also to be expected for the IResNet. To assess the uncertainty of the classification, we compute the mean metrics with standard deviation for each of the five validation splits in the crossvalidation and obtain an accurracy of $0.886\pm 0.003$ and F1-score of $0.472\pm 0.005$.
We demonstrate in our supplementary materials that the predicted classes are also physically consistent with the expectations of the respective WMO cloud types.

\begin{table*}[]
    \caption{Fractions of the cloud classes for pixel-wise classification with prediction accuracy and F1 score for the supervised part of the data. CloudSat labels are for $~21\cdot10^6$ labelled pixels included in \textsc{Cumulo}, predictions are for $~800\cdot 10^6$ pixels. Scores are averages for the $5$ validation splits.}
    \centering
    \begin{tabular}{|l|c|c|c|c|c|c|c|c|}
                \hline
                & Ci & As & Ac & St & Sc & Cu & Ns & Dc \\
         \hline
        CloudSat fraction & 0.259    & 0.132 &  0.112  &  0.021   & 0.313  & 0.065&    0.082 &  0.015 \\
        Predicted fraction & 0.154 & 0.10 & 0.180 & 0.027 & 0.353 & 0.014 & 0.134& 0.041\\
        Prediction accuracy &0.85 & 0.84& 0.88 & 0.95 & 0.78 & 0.91 & 0.90 & 0.98  \\
        Prediction F1-Score &0.64 & 0.40 & 0.44 & 0.21 &0.66 & 0.46 & 0.48 & 0.48\\
        \hline
    \end{tabular}
    \label{tab:histo}
\end{table*}

\subsection{Cloud class distributions at coarse resolution\label{sec:res_reg}}
The RF is expensive to train on large datasets such as the year-long, high-resolution \textsc{Cumulo} dataset. Because of these computational constraints, we train the Random Forests on a subset of the labelled data of about 25\% the size of the complete dataset. Tests have shown that the errors stabilize when using even fewer training data. The mean errors and R2 scores for the different settings are summarized in Table~\ref{tab:deviation}. Since with smaller grid cell sizes, more cells containing only ``undetermined'' pixels are excluded, and the relative amount of cloudy pixels increases, which is why we see larger mean absolute errors for small grid cells. Using the median, however, we see better performance for smaller cells. The R2-score increases with grid cell size, most likely due to the decreasing variance caused by averaging over more pixels. The performance is therefore judged not to be strongly dependent on the grid box size.
We also use joint densities of predicted and ground truth cloud fractions of the test split as a performance indicator. Fig.~\ref{fig:joint100} shows these for a grid cell size of $(100\;\rm{km})^2$ using the optimal set of features (see section~\ref{sec:input}). The joint density plot displays the concentration of samples in the truth/prediction space, and along the $x$- and $y$-axis the marginal distributions of the true and predicted fractions, respectively.

For both cloud classes in Fig.~\ref{fig:joint100}, there is a clear correlation between the ground truth and the predictions with a Pearson correlation of $c_P=0.96$ for Ns and $c_P=0.89$ for Ac. Many predictions are, however, far off the target: Fig.~\ref{fig:Ns_100_9697} shows several hundred samples with a predicted Ns fraction of about $0.2$ where the true fraction is close to $1$. For this specific example, this is a small fraction ($\mathcal{O}(0.001\%)$) of the total number of grid cells, but it shows that the predictions can differ strongly from the true values in a non-negligible number of cases. This deviation is a manifestation of ambiguity between different cloud states, likely caused by noise generated by the averaging of the features. Furthermore, this is an example of the predictions favoring low cloud fractions, as the ``undetermined'' class is prevalent in the training data. As shown in Fig.~\ref{fig:Ac_100_9697}, large altocumulus (Ac) fractions $(>0.9)$ are underestimated by the RF, but the deviation in this region remains largely below $0.1$, as indicated by the dashed line. Most of the samples are, however, still contained within the $\Delta=0.1$ range (magenta dashed lines). For fractions larger than $0.2$, samples deviating by more than a factor of 2 (outside black lines) are rare. For fractions smaller than $0.2$ (bottom left corner), deviations by a factor of more than 2 occur frequently, indicating difficulty in correctly predicting small fractions. Note that such predictions contribute significantly to the relative error, but have a negligible effect on the absolute error. We construct a random baseline by sampling from the class distributions in the IResNet predictions. We find that the mean absolute deviation is larger for the random baseline by roughly a factor of five, indicating that the regression model outperforms the random baseline.

\begin{table*}
    \caption{Results of the regression models for different grid box sizes. (1): Trained using a default set of features. (2): Using \textit{cwp} and \textit{cer}, not separated into ice and liquid, in addition to \textit{cod, ptop, tsurf}. (3): Using  (\textit{cwp, cer, ptop,  tsurf}) as features.}
    \centering
    \begin{tabular}{|c|c|c|c|c|c|c|}
    \hline
        \multicolumn{1}{|p{1cm}|}{Grid cell size} & \multicolumn{1}{p{1cm}|}{Mean abs. error} &  \multicolumn{1}{p{1cm}|}{Median error} & \multicolumn{1}{p{1cm}|}{Median relative error} & R2-Score &
        \multicolumn{1}{p{1cm}|}{Random MAE}\\
        \hline
        	3\;km\;$^{(1)}$ &0.042		&0.0018	&18.9\%	&0.816	& 0.193 \\
            10\;km\;$^{(1)}$&0.036		&0.002	&33.9 \%	&0.836	&0.178	\\
            20\;km\;$^{(1)}$&	0.033	&0.0025	&41.3\%	&0.840	&0.169	\\
            100\;km\;$^{(1)}$&0.027		&0.0038	&52.6\%	&0.845	& 0.151	\\	
            200\;km\;$^{(1)}$& 0.025		&0.0045	&54.2\%	&0.859	& 0.143	\\
            100\;km\;$^{(2)}$&0.028	&0.0041	&55.4\%	&	0.837&0.151	\\
            100\;km\;$^{(3)}$&0.033	& 		0.0043&60.4\%	&	0.755& 0.151\\
        \hline
     
    \end{tabular}
    \label{tab:deviation}
\end{table*}

In the supplementary materials, we show that our method reproduces physically meaningful cloud class distributions. Specifically, we show that the features resulting in a prediction of a certain cloud class are in line with those expected from the meteorological definition of the respective class.

\input{figs/fig6}

\input{figs/fig7}

As the variables available in typical ESM output can vary, not always matching our optimal set of features, we also determine which of these features are essential to achieve good performance. In addition to using the optimal set of features, we therefore also trained the model using different alternative sets, containing fewer features. Using the cloud top phase flag to distinguish between ice and liquid for some of our features (\textit{lwp, iwp}, \textit{cerl,ceri}) produces a small performance increase as the metrics indicate in Table~\ref{tab:deviation}. Comparing Fig.~\ref{fig:Ac_100_9697} and Fig.~\ref{fig:Ac_0126}, shows that the correlation between the true and predicted values becomes less pronounced when the information about the thermodynamic phase is removed. Further ablation studies reveal, that using \textit{cod} and \textit{ptop} is critical for the RF performance, but these variables are infrequently contained in ESM standard output. An example is shown using features without \textit{cod}, where Table~\ref{tab:deviation} shows a significant decrease in the R2-score. The effect on the joint density displayed in Fig.~\ref{fig:Ac_026_bad} is visible as predicted fractions being skewed towards smaller values. Predicted fractions above $0.8$ are very rare and the joint density seems to be shifted towards the lower black line, corresponding to half the true value. We provide further information on the feature importance in the supplementary materials.

\subsection{Validation\label{sec:res_val}}
\begin{figure*}
	\centering
    \includegraphics[width=\linewidth]{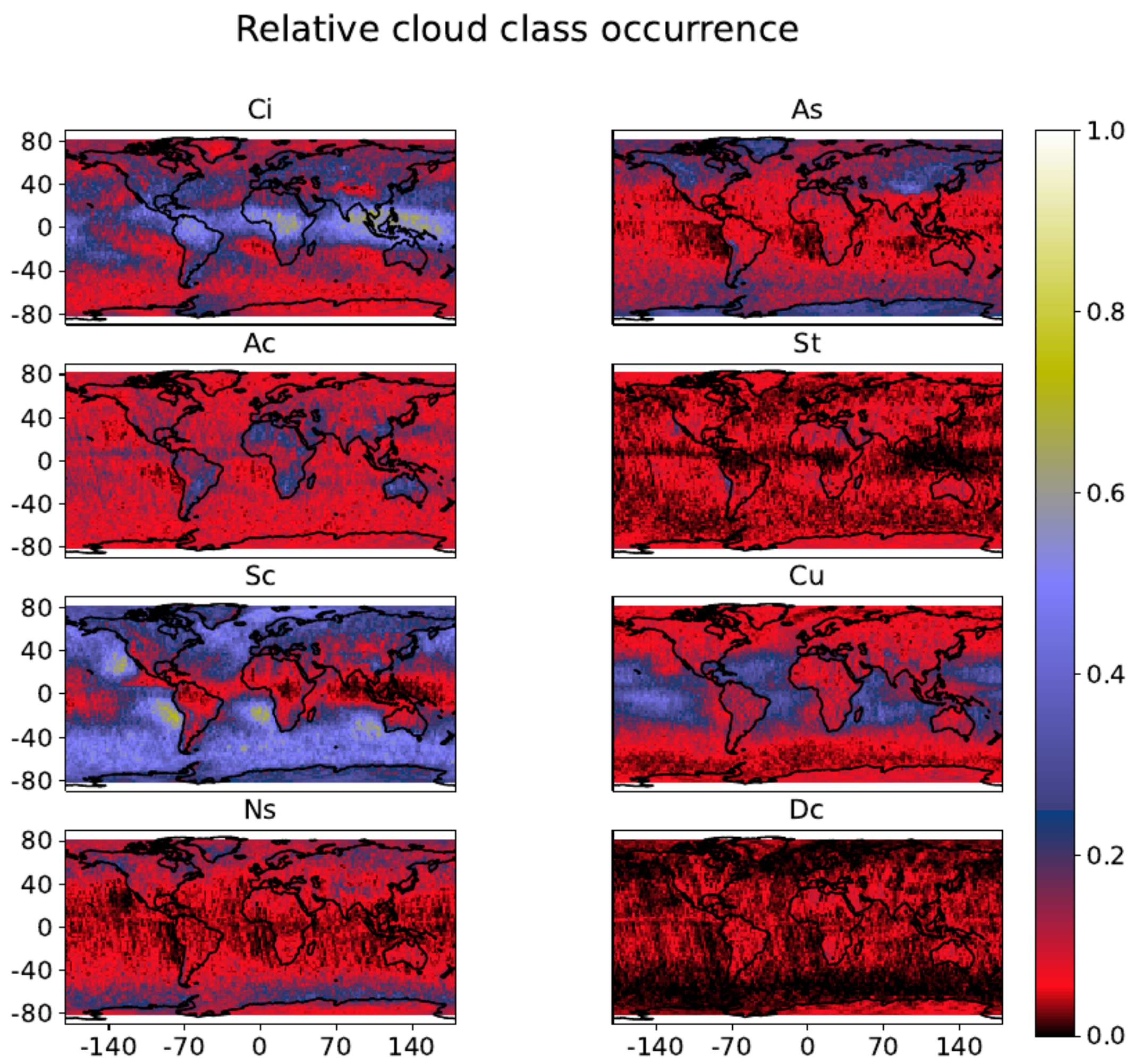}
    \caption{Relative occurrence per class from CloudSat measurements (year 2008) from the 2B-CLDCLASS-LIDAR product per $2^{\circ}\times 2^{\circ}$ grid cell.}
    \label{fig:CloudSat}
\end{figure*}

To assess the generalization performance of the method, we compare predictions of the class distributions on ESA CCI data to the classes from the CloudSat CLDCLASS-LIDAR product. We use the class labels from CloudSat for the year 2008. Again, the 3-dim CloudSat data are aggregated into two dimensions by using the most common cloud class within each vertical column as a representative cloud class. The labels provided by CloudSat for individual orbits are sparse, but using a whole year of measurements provides enough samples per location to compare to the predicted distributions. Fig.~\ref{fig:CloudSat} shows the sparsity of CloudSat labels, even though the data have been aggregated to grid cells of $2^{\circ}\times2^{\circ}$. Consequently, clear regional differences are not visible for all cloud classes, but Ci, Sc, and Cu show distinct areas of frequent occurrence. For example, Sc clouds are frequently detected in the subtropical subsidence regions off the west coasts of the continents, Ci clouds are frequent in the deep Tropics, Cu is found frequently over the tropical and subtropical oceans away from the stratocumulus decks. 

In the following, for a better comparison of the CloudSat ground truth and the predictions on the ESA CCI data, we exclude the ``undetermined'' predictions such that the cumulative fraction of all eight cloud classes equals one in each cell. The reported fractions are therefore a relative measure and independent of the total cloud amount in each grid cell.

The \textsc{Cumulo} and the ESA CCI data are provided on different grids. Because of averaging, same-size grid cells are not needed and thus no interpolative re-gridding is applied. We compare predictions obtained with the same model on differently sized ESA CCI grid cells in the supplements to this paper. Our results show that for a model trained on large $(100\;\rm{km})^2$ grid cells, there is no qualitative difference in the predictions for different grid sizes.  However, some features become more pronounced when smaller grid cells are used during prediction. The maximum grid cell size for a reasonable application to ESM output, therefore, depends on the region or processes of interest.

\begin{figure*}
    \centering
    \includegraphics[width=\linewidth]{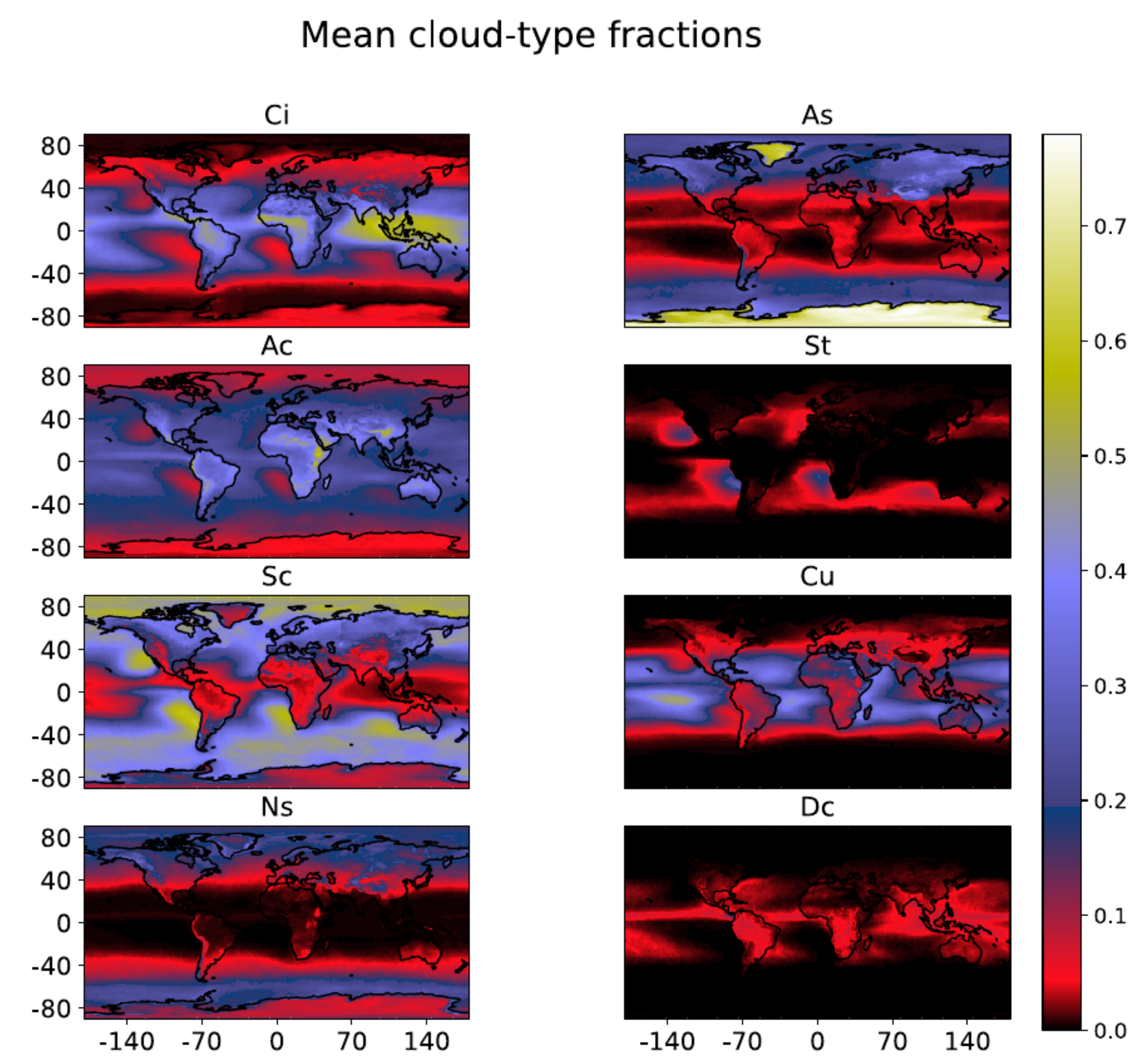}
\caption{Average class fractions for the predictions on coarse-grained ESA CCI data. The  RF was trained on $(10\;\rm{km})^2$ grid cells and applied to $10\times 10 $ pixel grid cells. The results are projected onto a $1^{\circ}\times 1^{\circ}$ grid. Many classes show similarities to the distributions in the CloudSat data (Fig.~\ref{fig:CloudSat}), even though the location is not used as a feature.}
    \label{fig:allctypes_higres}
\end{figure*}

Fig.~\ref{fig:allctypes_higres} shows the predictions on the ESA CCI data using the RF trained on grid cells of $(10\;\rm{km})^2$ and applied $10 \times 10$ pixel grid cells. Different classes occur in distinct patterns and the Sc class dominates in the predictions, while St and Dc occur very rarely.

\begin{figure*}
    \centering
    \includegraphics[width=0.9\linewidth]{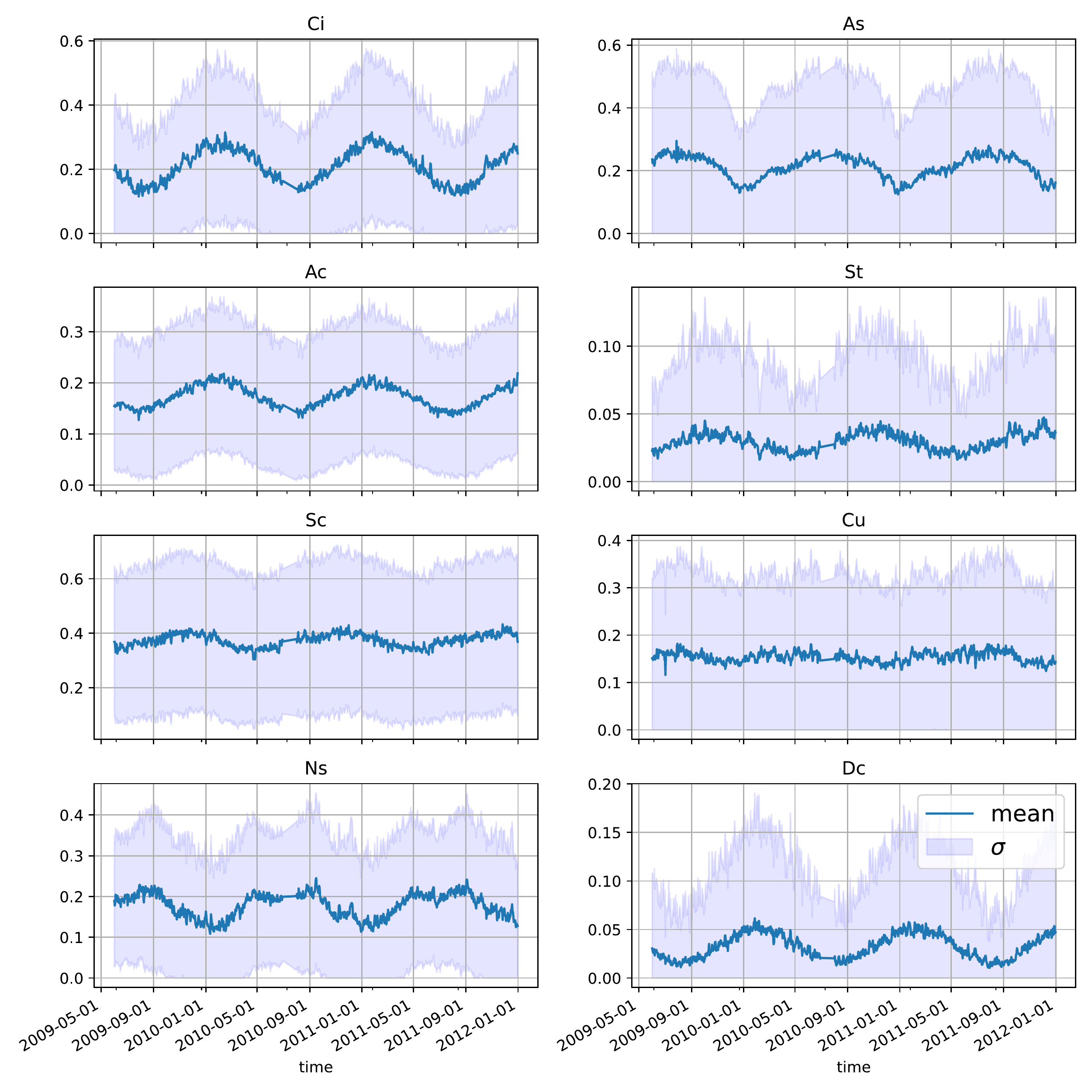}
    \caption{Time series of mean predicted class fractions from 1 June 2009 to 31 December 2011. The per-class mean is computed for all locations in the southern hemisphere where at least one instance of the respective cloud class fraction is within the 90th percentile of all predictions of this class. Note that July 2010 has been excluded due to faulty data.}
    \label{fig:temporal}
\end{figure*}

Fig.~\ref{fig:temporal} shows the time series of the class fractions averaged over grid cells in the Southern Hemisphere for which the respective cloud class can vary strongly. We define this by selecting grid cells for each class where at some time the class fraction is especially large (90th percentile).
Using this method of analysis, almost all predicted classes show a seasonal cycle. Only for the classes Cu and Sc such a cycle is not visible. The Ns and As classes are predicted at higher fractions in the cold months. In contrast to Ns and As, Ac and St have higher fractions in summer. The CloudSat ground truth is too sparse to similarly assess seasonal cycles and enable a direct comparison.

\subsection{Uncertainty estimate}
For the ESM evaluation, deviations introduced by the data need to be separable from those caused by the evaluation method.

\begin{figure*}
    \centering
    \begin{subfigure}{0.49\textwidth}
        \centering
        \includegraphics[width=\linewidth]{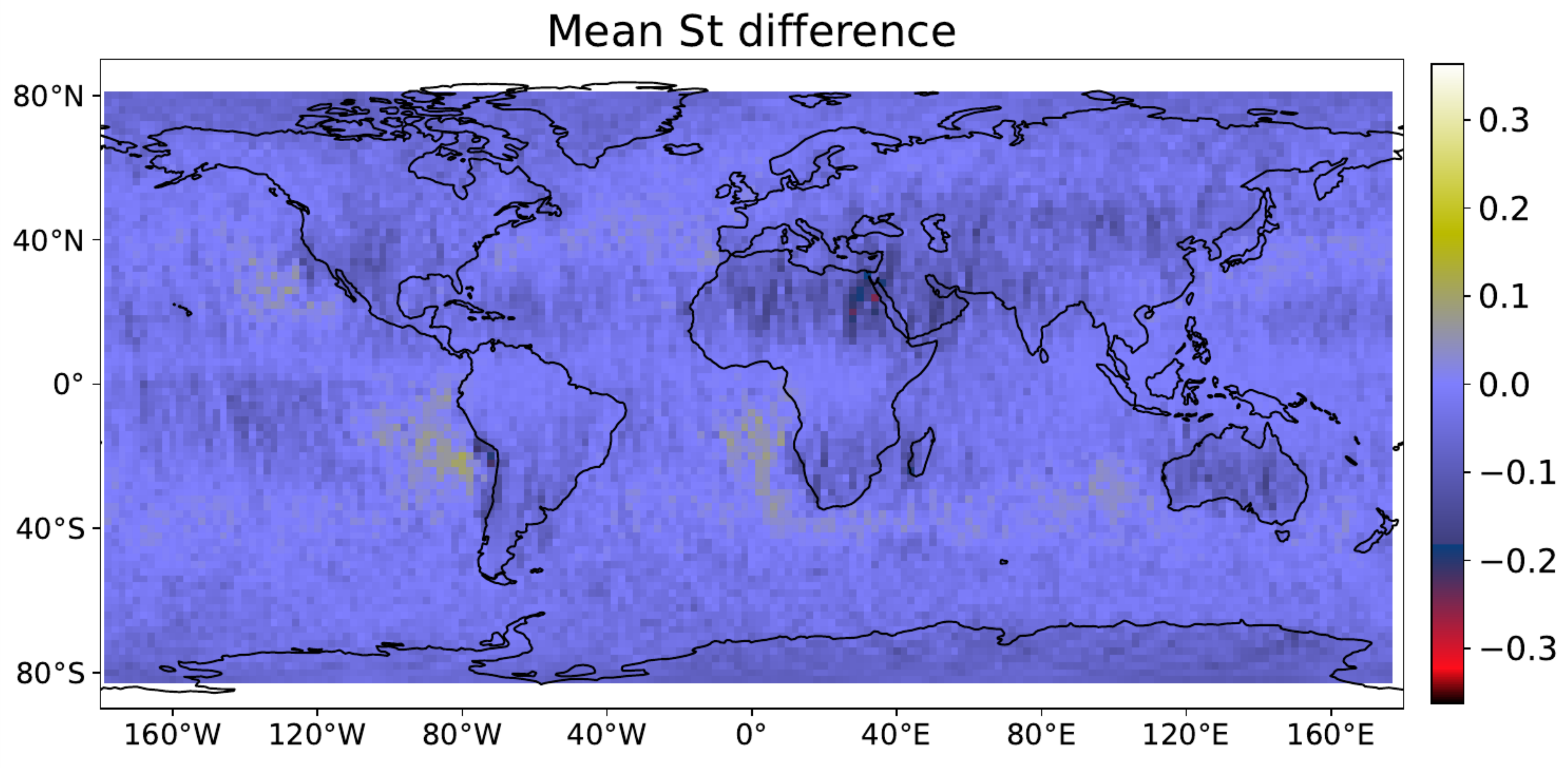}
        \subcaption{}
        \label{fig:DiffSt}
    \end{subfigure}
    \begin{subfigure}{0.49\textwidth}
        \centering
        \includegraphics[width=\linewidth]{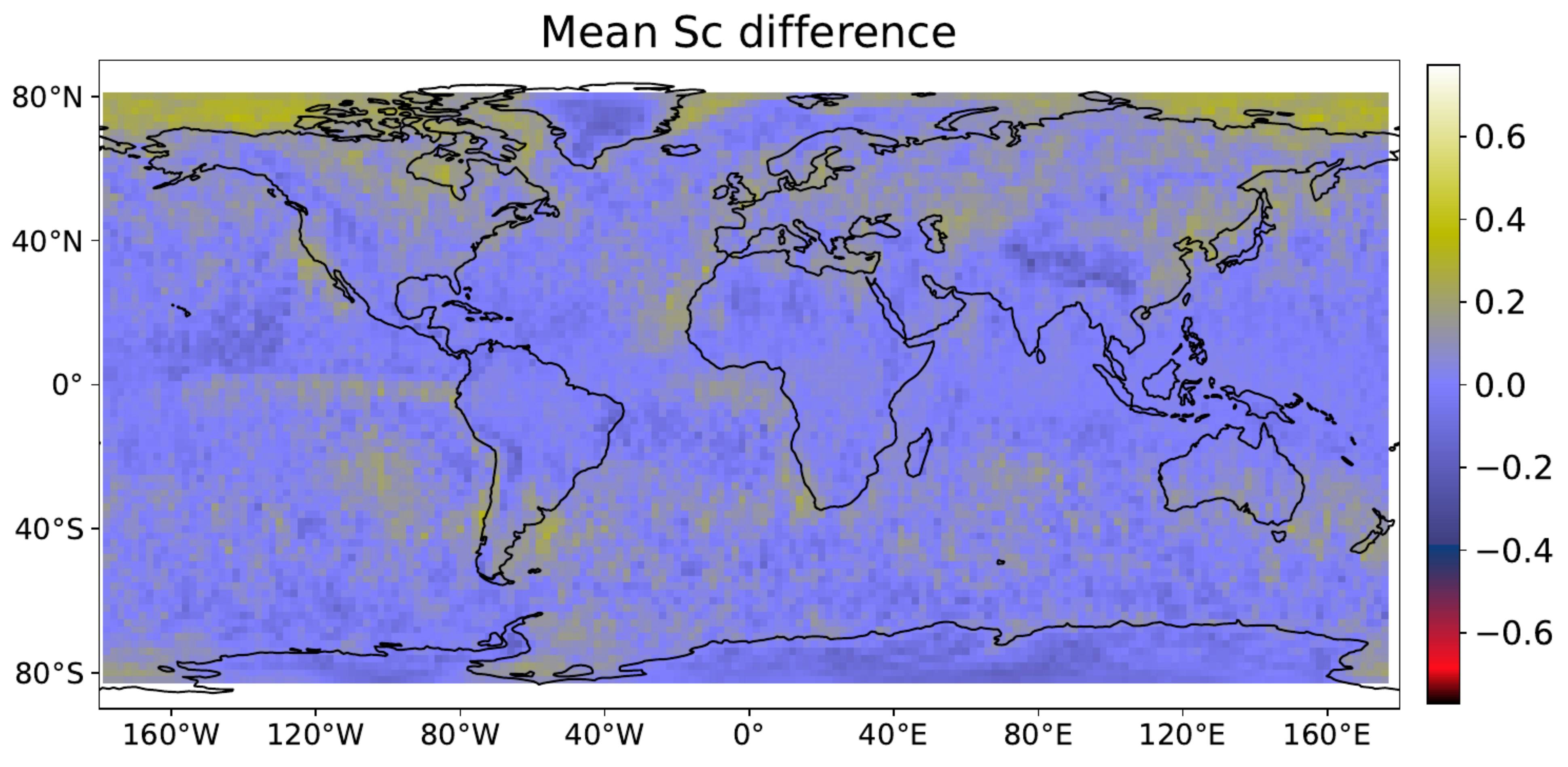}
        \subcaption{}
        \label{fig:DiffSc}
    \end{subfigure}
    \caption{Difference between the mean predicted fractions and CloudSat per $2^{\circ}\times2^{\circ}$ grid cell for the relative amount of the classes with lowest/highest correlation, St (\ref{fig:DiffSt},$c_P=0.18$)/Sc (\ref{fig:DiffSc},$c_P=0.88$). The color map is normalized to the range $[-m,m]$, where $m$ is the maximum value for the class across both (CloudSat, predictions) distributions. Predictions obtained from RF trained on $(100\; km)^2$ data and applied on $100\times 100$ ESA CCI pixels.}
    \label{fig:diff}
\end{figure*}

As an uncertainty estimate for the consecutive application of both machine-learning methods, we compute the Pearson correlation and difference between the predictions and the CloudSat labels. For this purpose, we bin the relative amount of each class to grid cells of $2^{\circ}\times2^{\circ}$ size for both datasets. As an example, differences for the cloud types with highest (Sc) and lowest (St) correlation are shown in Fig.~\ref{fig:diff}. The difference increases with the fraction of occurrence of each class (as displayed in Fig.~\ref{fig:allctypes_higres}). Note that this is only a rough measure of accuracy as the two datasets differ in temporal and spatial resolution. Additionally, an exact match cannot be expected as the CloudSat data cover the year 2008, while the ESA CCI data cover the period June 2009 to the end of 2011. The mean within-class correlation is $0.65$. Table~\ref{tab:cloudsatcomp} shows the mean fractions of the classes in the predictions and the CloudSat data. The predictions here are comparable to the pixel-wise predictions obtained using the IResNet (Table~\ref{tab:histo}). The most notable difference in the distribution is again the under-representation of Ci in the predictions relative to the CloudSat labels, which is caused by the under-representation in the predicted pixel-wise labels. Table~\ref{tab:cloudsatcomp} also shows the relative difference between the two distributions for grid cells showing a large class fraction in the predictions (90th percentile). For all classes, the magnitude of this deviation is below 50\%. This is also the range of relative deviation we found on the test split, leading to an overall estimate of the uncertainty of $50\%$.

\begin{table}[]
    \caption{Mean fraction of the predicted classes compared with the relative amounts of the classes in CloudSat. The last row shows the mean difference for pixels with predictions in the 90th percentile $\Delta_{90}$ relative to the mean $\mu_{90}$ of these predictions. Predictions are taken from a model trained on the default set of features using $(100\;\rm{km})^2$ and applied on 100$\times$100 pixel ESA CCI grid cells.}
    \label{tab:cloudsatcomp}
    \centering
    \begin{tabular}{l|c|c|c|c|c|c|c|c}
                \hline
                &Ci &As &Ac &St &Sc &Cu &Ns &Dc \\
         \hline
        Predict. & 0.13 & 0.14 & 0.19 & 0.01 & 0.31 & 0.10 & 0.10 & 0.02\\
        CloudSat & 0.20 & 0.13 & 0.11 & 0.05 & 0.27 & 0.12 & 0.09 & 0.03 \\
        $c_P$ & 0.87 & 0.80 & 0.60 & 0.18 & 0.88 & 0.84 & 0.83 & 0.36\\
        $\Delta_{90}/\mu_{90}$ & -29\% & 49\% & 49\% & 1\% &18\% & 14\% & 30\%& 39\%\\
        \hline
    \end{tabular}
\end{table}

\section{Summary\label{sec:summ}}  
We presented a method for the evaluation of clouds in coarse resolution data, employing the consecutive application of machine-learned classification and regression models. Using this method, information on clouds from high-resolution, three-dimensional CloudSat and CALIPSO lidar products is first added to passive sensor data from MODIS by using the \textsc{Cumulo} framework and then transferred to coarse resolution data. This approach offers a new perspective on statistical and process-oriented assessment of the performance of climate models by being able to analyze the model output in terms of different cloud classes and thereby distinguishing the driving mechanisms for the formation and evolution of different cloud types more clearly. This provides the potential to better understand and ultimately improve existing model deficiencies.

The pixel-wise classification has a high accuracy of at least $0.8$ for each class, with little variation across the validation splits. The relative amounts of predicted Cu and St, however, can differ by more than a factor of 2 compared to the CloudSat data used as ground truth. The predicted cloud classes show distinct physical properties that are consistent with the expected properties of the corresponding WMO cloud classes. 

The subsequent regression can reproduce consistent cloud class distributions on regional scales with mean errors being at least one order of magnitude smaller than the random baseline. Furthermore, the RF regression successfully generalizes to different data as could be shown using the ESA CCI data. The predicted global distributions of the individual cloud classes compare well with the CloudSat ground truth. This is evident when qualitatively comparing the distributions for each class as well as in the correlations and differences in areas of a high-class fraction. The correlation is larger than $0.6$ for all but 2 classes (St/Dc) and the relative difference in areas of a high-class fraction is smaller than $50\%$ for all classes. The spatio-temporal location of a sample is not used as an input for the machine-learning algorithms. Therefore, any predictions are solely based on the physical properties represented by the features. Yet, even small-scale regional characteristics of the CloudSat ground truth are similarly represented in the predictions using the ESA CCI data. Notable examples are a peak of Ci in the tropical warm pool region or an increased As fraction in the Himalayas. Additionally, the geographical means for all classes correlate positively with the respective relative occurrence in the CloudSat ground truth, with higher correlations for the classes with many available samples. 
We further showed that the regression model associates each class with specific feature values (see supplements). These values are consistent with the expected properties of the different WMO cloud types. 
Analysis of the effects of temporal averaging of the target data showed that the method works well with near-instantaneous data but cannot be applied to monthly averaged data (see supplements). 

Tests with multiple sets of input features have shown that information about the cloud height, cloud water content and optical thickness are essential for good performance, with information on the thermodynamic phase of the cloud providing additional robustness. In contrast, the horizontal resolution of the data the model is trained on seems to be less critical (see supplements). Models trained on different grid cell sizes show differences but no clear optimal resolution can be defined from these initial results. It is recommended, however, that predictions be performed using data at their highest available horizontal resolution as more features can be resolved.

Predictions on test data show clear correlations of ground truth and predictions (average $\Bar{c}_p =0.92$). Predictions on the ESA CCI data provide enough information to isolate individual features and processes. This suggests that this method can be successfully applied to any dataset of sufficient length and horizontal and temporal resolution to allow for statistically robust predictions. 
When applied to the test data, the median relative deviation was about $50\%$. Comparing the predictions with the raw labels from CloudSat we find similar values. Especially in regions where specific cloud types are predicted with a high frequency of occurrence, we find relative deviations mostly below $50\%$. Only the St class is consistently underestimated for which the pixel-wise classification already showed poor performance.

\section{Discussion\label{sec:Dis}}
The results from testing the regression model with unseen data show physical and temporal consistency of the results across all analyses. This is the primary goal of this method aiming to evaluate physical processes. We can therefore be confident that the results are meaningful, even though the results of the classification do not exactly reproduce the label distribution in the source data. The deviations in the amount of St predicted by the classifier can be at least partly explained by the relatively small amount of training samples and the similarity to the physical properties between St and Sc. We conclude that pixel-wise labelled data are therefore suitable as a basis for training a regression model which learns cloud class distributions on datasets with a horizontal resolution typical for climate model scales. Generally, our results suggest that the method is, therefore, suitable for a process-oriented assessment of clouds simulated by climate models. Using the predicted distributions this can be performed in the space of cloud classes, providing several advantages. Firstly, a layer of subjective interpretation is removed by being able to analyze the results in terms of cloud classes clearly defined by the underlying classification algorithm (CloudSat, \cite{PCICD2019}). Because we are using the commonly used WMO classes, the resulting cloud class distributions can then be analyzed and interpreted in a process-based manner as the key processes driving formation and evolution differ between the cloud classes. This greatly simplifies the analysis and evaluation of selected physical properties related to cloud processes in the climate models. Secondly, as the deep learning algorithm learns from high-resolution 3-dimensional data, the climate models are implicitly analyzed in a horizontally super-resolved manner which also takes into account information about the vertical structure of clouds, i.e. learning from a combination of 2- and 3-dimensional data can potentially take advantage of information from the vertical that would not be included in the cloud top view. That the method can resolve phenomena on regional and seasonal scales provides the opportunity to identify spatio-temporal areas in which clouds are not correctly represented. This could for example be done for the low-level clouds found in the subtropics west of the continents, investigating their horizontal extent, their dependence on feature values, and their temporal evolution. 
However, due to the nature of the multi-stage process, some limitations apply: by building the regression on 2-dimensional, spatially averaged source data it is hard to make correct predictions on individual grid cells. This results in several samples for which the predicted cloud fraction differs by a factor of 2 or more. Additionally, the under-representation of the Cu class and the limited accuracy of the St class show that this method can still be improved. This probably stems at least partly from the CloudSat ground truth itself, as the CloudSat algorithm has trouble distinguishing between St and Sc. We, therefore, recommend combining these classes when applying these or similar methods. Some features of the predicted cloud distribution such as, for example, the high fractions of Ns along the Antarctic coast, might be amplified or hidden by noisy satellite retrievals. Especially in high latitudes, clouds can be challenging to characterize with passive sensors like MODIS. 
Our ML models do not provide satisfactory results when applied to temporally averaged data because they are trained on instantaneous measurements. Using geostationary data available e.g. every 30 minutes (GOES satellite, \cite{Walther2013}) for the pixel-wise classification instead of MODIS data available only twice a day might improve the results. Such an approach has been applied to other atmospheric variables like convection and rainfall \cite{Lee2021,Gorooh2020}. The physical properties of the predicted clouds could then be safely averaged over time due to the high and consistent temporal resolution of the data allowing the regression model to train on data more comparable to typical ESMs output. However, the processes to be evaluated with our approach are not resolved at large temporal scales. This contributes to the poor regression performance for monthly mean data and will still be an issue when the RF is trained on temporally averaged data. In turn, this means that this method is suitable to detect model deficiencies relatively quickly in contrast to using climatological means from long-term simulations. This is because we would expect an inaccurate representation of the global and regional cloud distributions to be already detectable with model output available for less than a year.

The consecutive application of two machine learning steps makes it difficult to quantitatively estimate the propagated errors. Even though the error of the classification and regression can be individually estimated using test splits, the combined impact of these errors is not clear. The small variations of the different IResNets used in the crossvalidation suggest high confidence of the networks in their predictions, but errors or uncertainties can not be propagated through an RF. Error estimation is further complicated by possible inconsistencies in the CloudSat classification algorithm, i.e. clouds not necessarily being labelled the same way a human expert would. An example of this is the difficulties in differentiating between St and Sc. We do not, however, see any specific inconsistencies in the physical properties or the regional distributions of the predictions, suggesting that the propagated uncertainties are reasonable. We estimate the uncertainty of at least $50\%$ to what would be reported by CloudSat for individual predictions can be expected. However, in large datasets, the method can identify individual regions of high occurrence for a class. The absence or underestimation of such phenomena in the global cloud distribution would be signs of possible ESM deficiencies. Even for classes for which limited training data are available (Cu,St,Dc), we find that the predictions are self-consistent. This is apparent in the characteristic feature values, which are distinct for each class and do not vary regionally (see supplements). The regional distributions of the classes are attributable to the predominant atmospheric conditions. For example, the Dc class occurs more frequently near the equator, Cu is found predominantly in tropical and subtropical regions over the ocean, St is mostly west off the continents in the subtropical subsidence regions. Both, Cu and St are as expected low-level clouds with low cloud top heights.

In terms of implementation, this method can be applied to new datasets quickly and does not require individual implementation for each model. We would like to note that many of the variables needed for this method are typically part of the standard output of climate models so the main requirement would be to provide instantaneous or near-instantaneous values, i.e. model output not averaged over longer time intervals. We would therefore like to encourage the modeling community to provide such an output e.g. in future model intercomparison projects such as CMIP7. We especially recommend adding the cloud optical thickness, and the effective particle radii for liquid and ice particles as instantaneous 2-dimensional variables to the CMIP7 data request. 
Future improvement of the method could include replacing the RF as a regression model. The most significant advantage of the RF is the use of the bagging process during training, which helps to generalize well to unseen data. However, the size of the RF scaling with the size of the dataset and batched training is not straightforward requiring a high degree of subsampling to make training computationally feasible. Therefore, as noted previously, this required us to discard data. However, once trained, applying the RF to batched data for predictions is possible and fast, making it suitable for practical application. A CNN could be a reasonable replacement for the RF due to the resemblance of the data to images. First attempts to replace the RF with a CNN, however, did not yield satisfactory results independent of specific architecture, with the network's loss not converging. Additionally, implementation and training of the RF are much simpler than that of a CNN, which makes the RF more suitable in practice.
Also, in this study, the CloudSat cloud classes in the source data are aggregated in the vertical dimension by assigning the most common class in the cloud column to the respective pixel. Even though this provides an implicit resolution of vertical features, a full classification in three dimensions would be a clear improvement. An improved representation of the vertical cloud structure might be obtained by using a more sophisticated aggregation algorithm. Taking into account the physical properties of the observed pixel in each vertical column might lead to more representative ground truth.
\newpage

\small 
\printbibliography 
\normalsize

\newpage

\include{supp}

\iftrue

\clearpage

\end{document}

%% file: figs/fig6.tex
\begin{figure*}
    \centering
    \begin{subfigure}{0.45\linewidth}
        \centering
        \includegraphics[width=\linewidth]{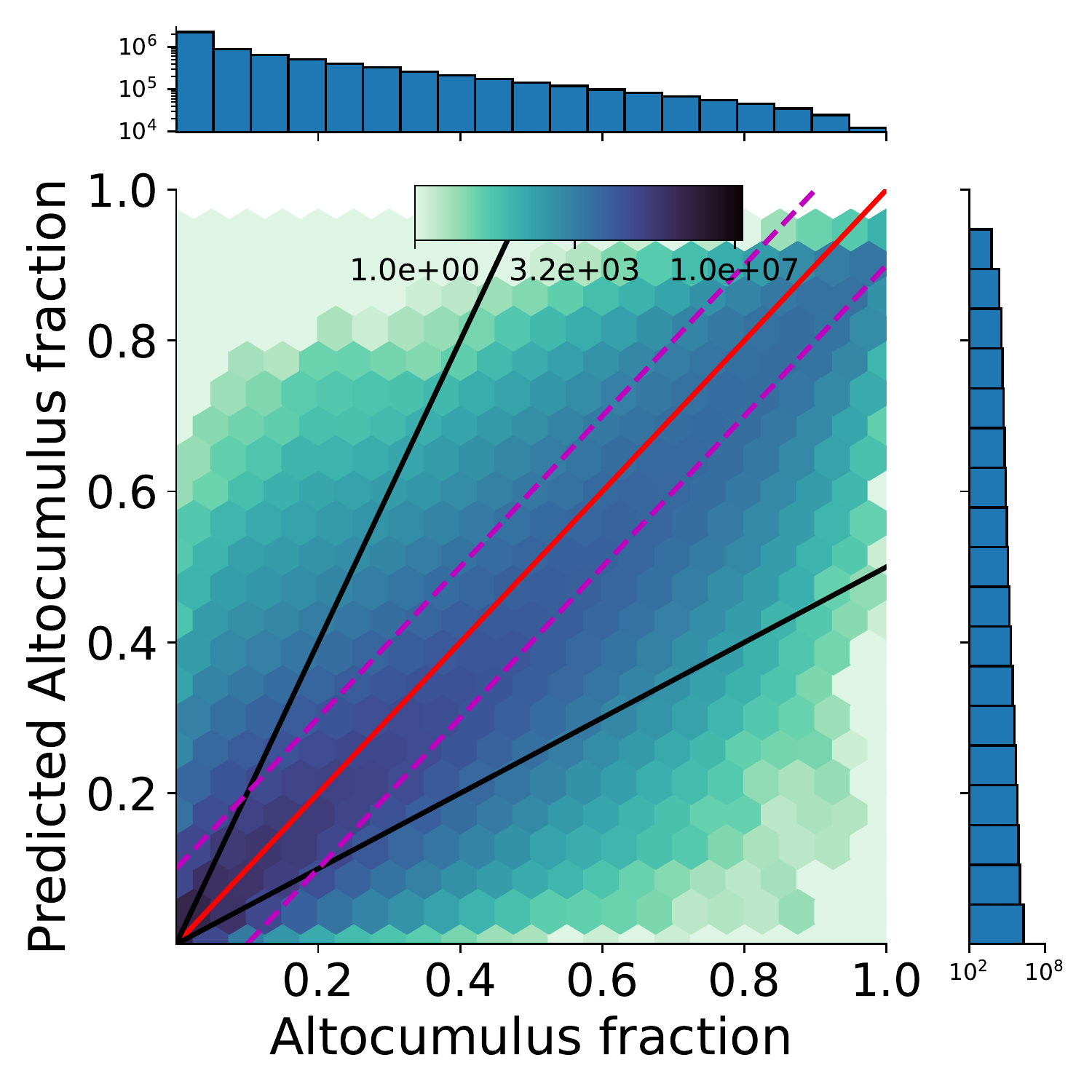}
        \subcaption{\label{fig:Ac_100_9697}}
    \end{subfigure}
    \begin{subfigure}{0.45\linewidth}
        \centering
        \includegraphics[width=\linewidth]{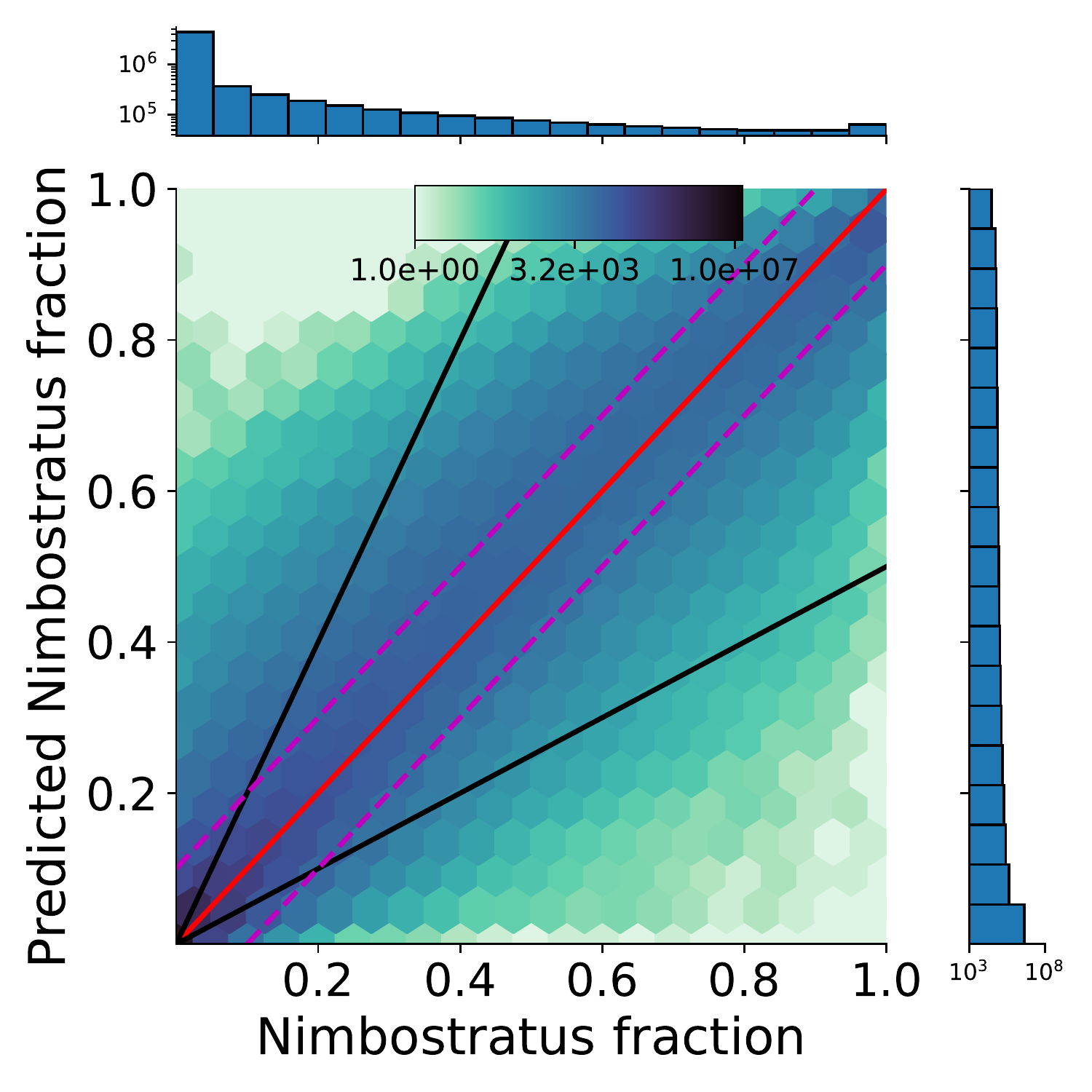}
        \subcaption{\label{fig:Ns_100_9697}}
    \end{subfigure}
    \caption{Joint density of the predicted and true Ac (left) and Ns (right) fractions from the \textsc{Cumulo} test set for a grid cell size of $(100\;\rm{km})^2$, using the optimal set of features (Section~\ref{sec:input}). The color scale and the marginal histograms are logarithmic. The red line indicates the line of perfect correlation. The area between the dashed magenta lines indicate a deviation between ground truth and prediction of less than $0.1$ and the area between the black lines indicates a deviation by less than a factor of two in either direction.}
    \label{fig:joint100}
\end{figure*}

%% file: figs/fig7.tex
\begin{figure*}
    \centering
    \begin{subfigure}{0.45\linewidth}
        \centering
        \includegraphics[width=\linewidth]{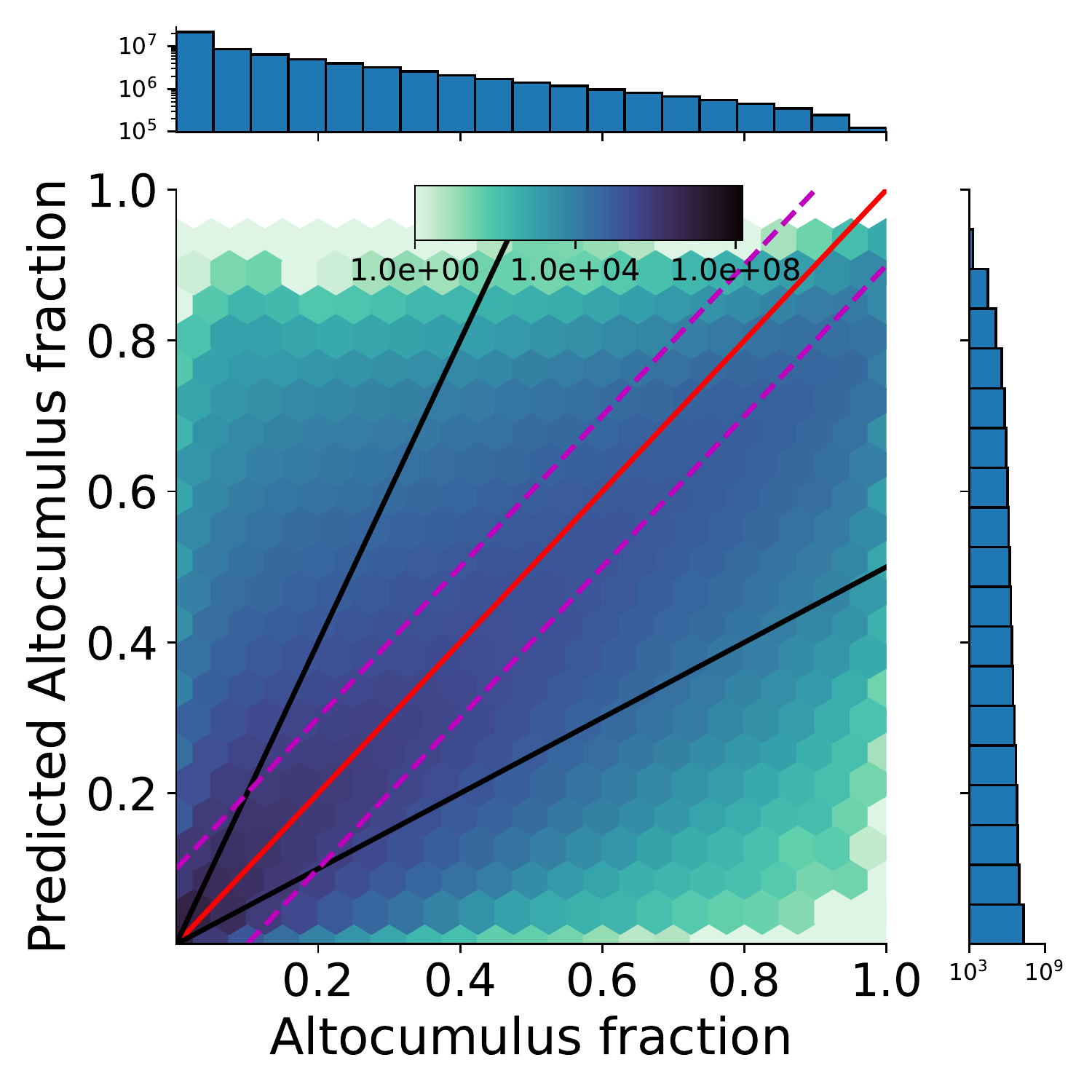}
        \subcaption{}
        \label{fig:Ac_0126}
    \end{subfigure}
    \begin{subfigure}{0.45\linewidth}
        \centering
        \includegraphics[width=\linewidth]{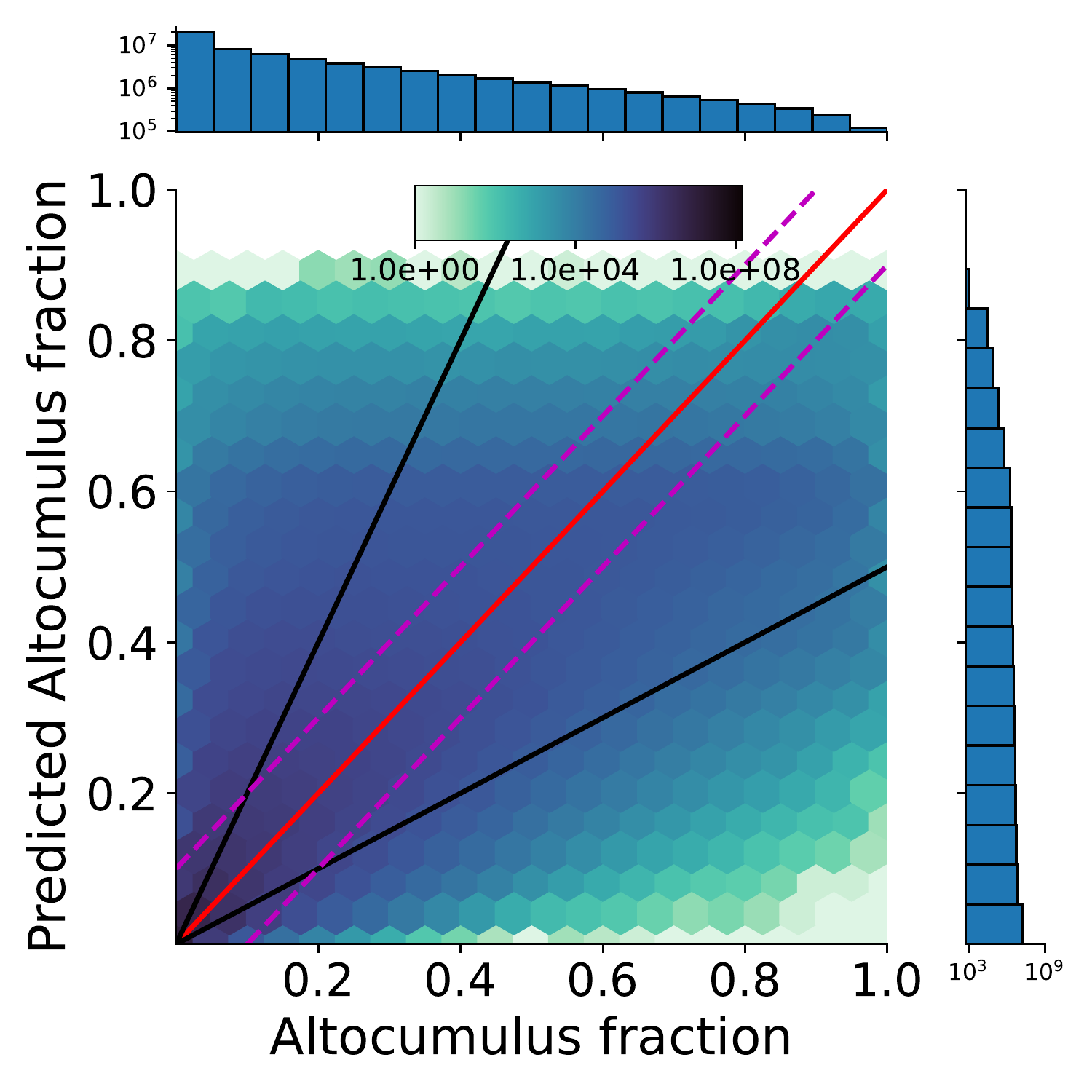}
        \subcaption{}
        \label{fig:Ac_026_bad}
    \end{subfigure}
    \caption{Results for a model trained using features without liquid/ice distinction (\textit{cwp, cer, cod, ptop, tsurf}) (\ref{fig:Ac_0126}) and a model trained also without \textit{cod} (\ref{fig:Ac_026_bad}), for comparison with Figure \ref{fig:joint100}. }
    \label{fig:joint100_phasecompare}
\end{figure*}

%% file: supp.tex
\begin{center}
\Large{Supporting information for the manuscript} \\
\textbf{``Machine-learned cloud classes from satellite data for process-oriented climate model evaluation''}\\[3ex]
\end{center}

Arndt Kaps\orcidlink{0000-0002-5368-5950},
Axel Lauer\orcidlink{0000-0002-9270-1044},
Gustau Camps-Valls\orcidlink{0000-0003-1683-2138},~\IEEEmembership{Fellow,~IEEE,}
Pierre Gentine\orcidlink{0000-0002-0845-8345},
Luis Gómez-Chova\orcidlink{0000-0003-3924-1269},~\IEEEmembership{Senior Member,~IEEE,}
Veronika Eyring\orcidlink{0000-0002-6887-4885}
\setcounter{page}{1}
\setcounter{figure}{0}
\setcounter{table}{0}
\setcounter{section}{0}
\setcounter{tocdepth}{1}
\renewcommand{\thefigure}{S\arabic{figure}}
\renewcommand{\thetable}{S\arabic{table}}
\renewcommand{\thesection}{S\arabic{section}}

\section{Overview}
In addition to the results presented in the main body of our paper, we illustrate four more ways to evaluate our methods performance in this supplemental document. We reinforce the validity of the results by showing that the predicted classes used in both our machine learning models are physically consistent with expectations. Furthermore, we investigate the importance of the different features used as inputs to our models with different metrics. We also compare results obtained by using different horizontal coarse graining resolutions as well as temporal resolutions.

\section{Physical consistency}

\begin{figure*}
        \centering
        \includegraphics[width=\linewidth]{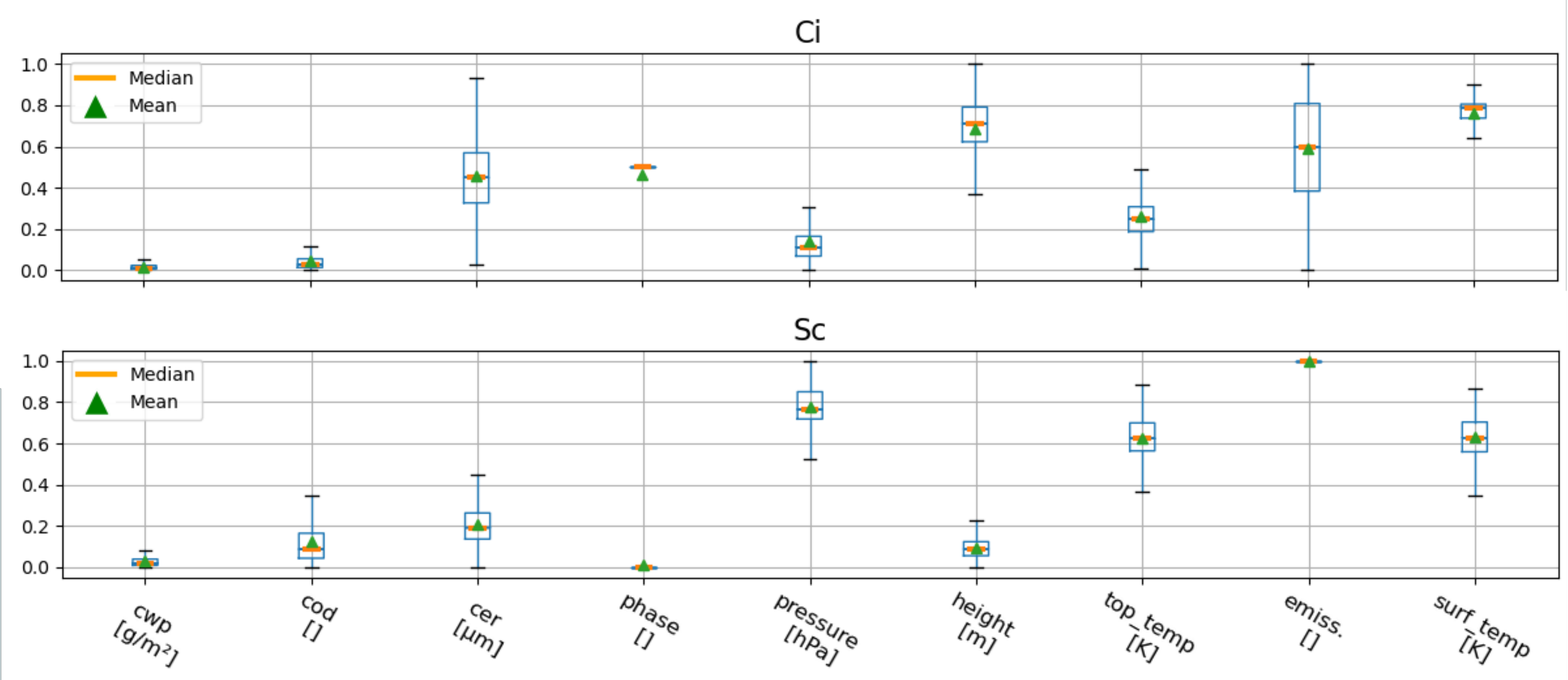}
    \caption{Distributions of the feature values for IResNet predictions of the two most common classes in the \textsc{Cumulo} classification, Ci (top) and Sc (bottom). The values are min/max scaled such that the feature values for all classes lie in the same range. Phase is a categorical feature: 0 for liquid, 0.5 for ice and 1 for undetermined. Boxes extend from lower to upper quartile, and whiskers extend from 10th to 90th percentile. Comparing the relative locations of the boxes between the classes allows assessment of the physical properties the IResNet associates with the respective classes. For example,  the cloud top pressure values for Ci are at the low end of the range, while the opposite is true for Sc.}
    \label{fig:vardist}
\end{figure*}

By analysing the features associated with each class we can show that the properties of the predicted classes are consistent with the expected properties from the meteorological definition of the respective cloud type (\href{https://cloudatlas.wmo.int/}{WMO Cloud Atlas}). In Fig.~\ref{fig:vardist}, we show the distribution of the features for the two most common classes Ci and Sc. The values for each feature are normalized to the range $[0,1]$, with zero and one corresponding to the minimum and maximum value observed across all classes. We find that the Ci class predicted by the IResNet has a higher cloud top and is exclusively flagged as ice. In addition, Ci clouds show a smaller water content and optical thickness. Similarly, we find that clouds are identified as Sc when a tile contains low-level clouds with a higher liquid than ice content and a higher optical thickness than Ci. Most other cloud classes also show consistent feature characteristics. The exception are the classes Sc and St, which the IResNet predicts for very similar ranges of the feature values. This is, however, in itself in line with the definitions of the cloud classes, as stratus and stratocumulus clouds have similar physical properties, leading to potential difficulties in the identification between the two classes.

\begin{figure*}
    \centering
    \begin{subfigure}{0.49\linewidth}
        \centering
        \includegraphics[width=\linewidth]{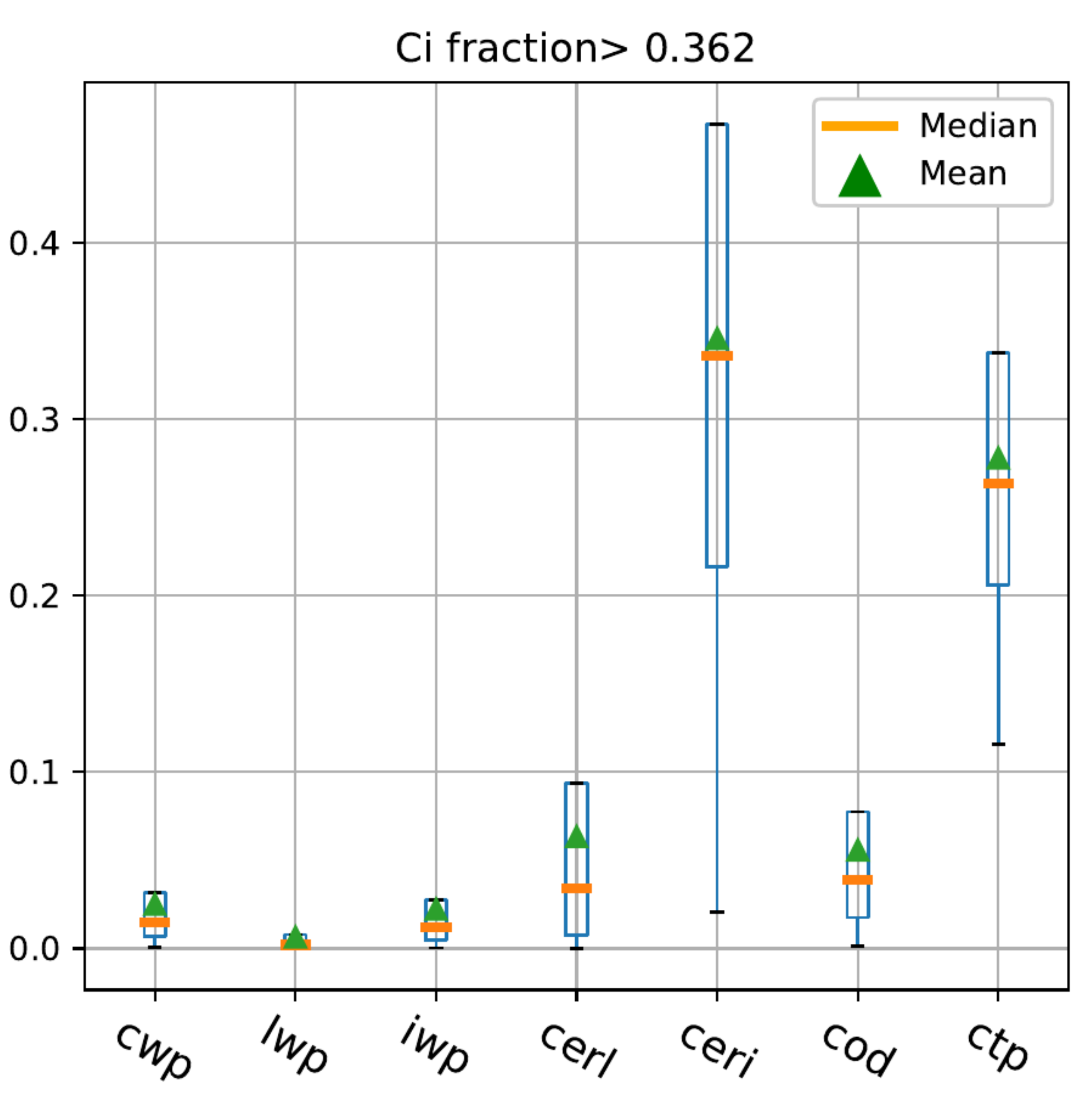}
        \label{fig:characteristics_Ci}
        \subcaption{}
    \end{subfigure}
    \begin{subfigure}{0.49\linewidth}
        \centering
        \includegraphics[width=\linewidth]{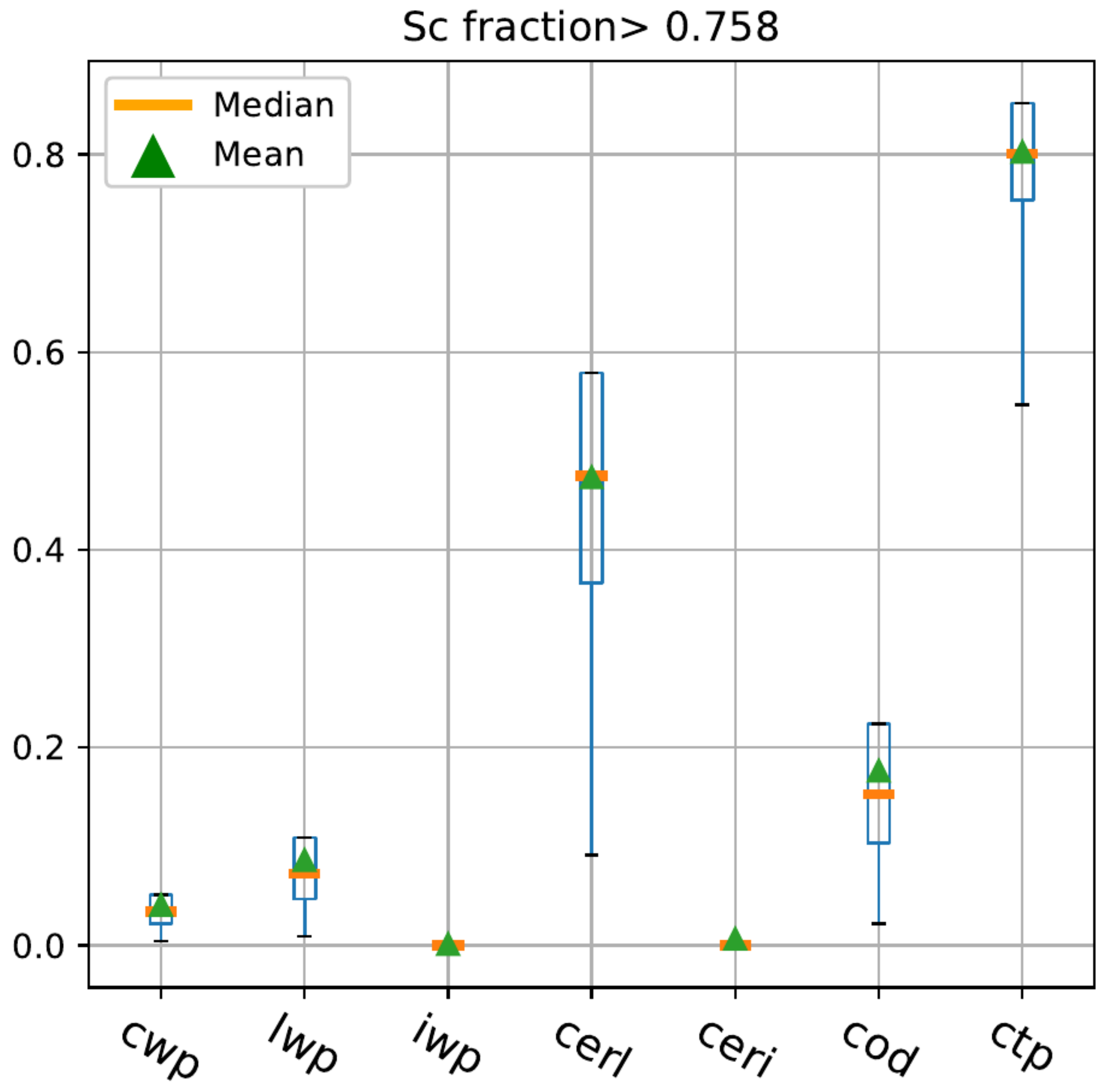}
        \label{fig:characteristics_Sc}
        \subcaption{}
    \end{subfigure}
    \caption{Distribution of the feature values of grid cells predicted by the RF to have a large fraction of Ci (left) or Sc (right) (90\% quantile, i.e. thresholds 0.362 for Ci and 0.758 for Sc; see also Fig.~\ref{fig:vardist}). The boxes indicate the ranges of the individual features which the RF associates with especially high occurrences of the respective class. Box ranges as in Fig.~\ref{fig:vardist}; values scaled such that 0 and 1 corresponds to the minimum and maximum values found across all classes, respectively.}
    \label{fig:characteristics}
\end{figure*}

The analysis of the physical consistency of the regression results requires additional steps. As most grid boxes include several classes, we cannot simply show the feature values associated with individual cloud classes like we did for the classification. Instead, we analyze the characteristic feature values for grid cells predicted to have an especially high fraction of a specific class. As an example, Fig.~\ref{fig:characteristics} shows the feature value distribution for grid cells predicted to have a high fraction of Ci or Sc classes. This allows for a comparison between the properties of the classes obtained in the pixel-wise classification (Fig.~\ref{fig:vardist}) and the class fractions of the grid boxes (Fig.~\ref{fig:characteristics}). Most prominent are the differences in liquid/ice particle radii between the two sets of grid cells. The respective values are consistent with cirrus being comprised of ice and stratocumulus consisting mostly of liquid particles. These plots also show that grid cells with a high fraction of Sc typically have higher cloud top pressures (i.e. lower cloud top heights) and contain more condensed water than grid cells with a high fraction of Ci. As expected, we find that the predictions of the regression model for high fractions of a cloud class are based on a similar feature values as the predictions on the pixel level. This shows that the regression model can indeed predict cloud classes from average states of large grid cells distinctly and that the physical basis used for the CloudSat classification is propagated throughout the individual steps of our method.

\begin{figure*}
	\centering
    \begin{subfigure}{0.49\linewidth}
    	\centering
        \includegraphics[width=\linewidth]{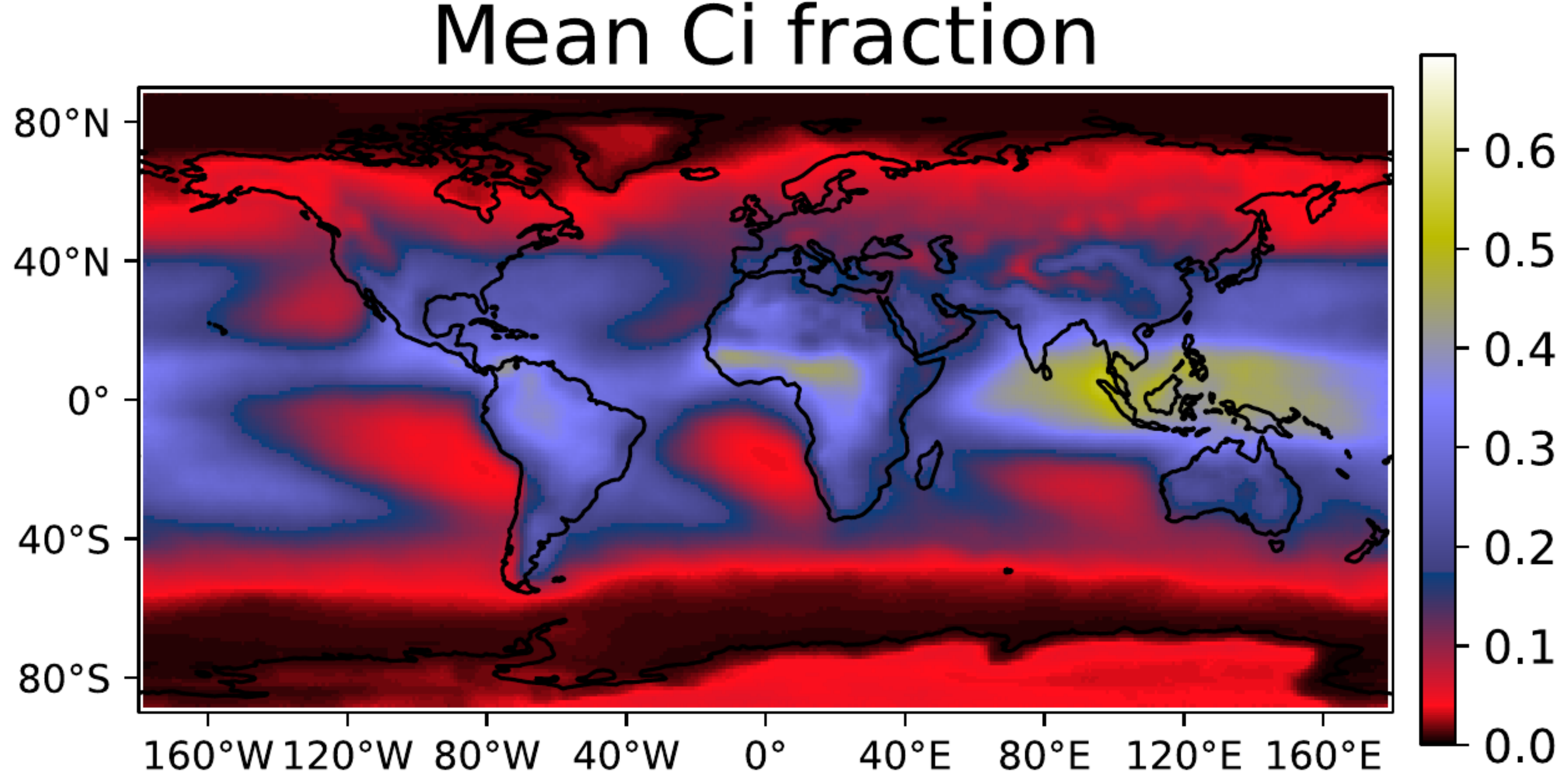}
        \subcaption{}
        \label{fig:Ci}
    \end{subfigure}
    \begin{subfigure}{0.49\linewidth}
    	\centering
        \includegraphics[width=\linewidth]{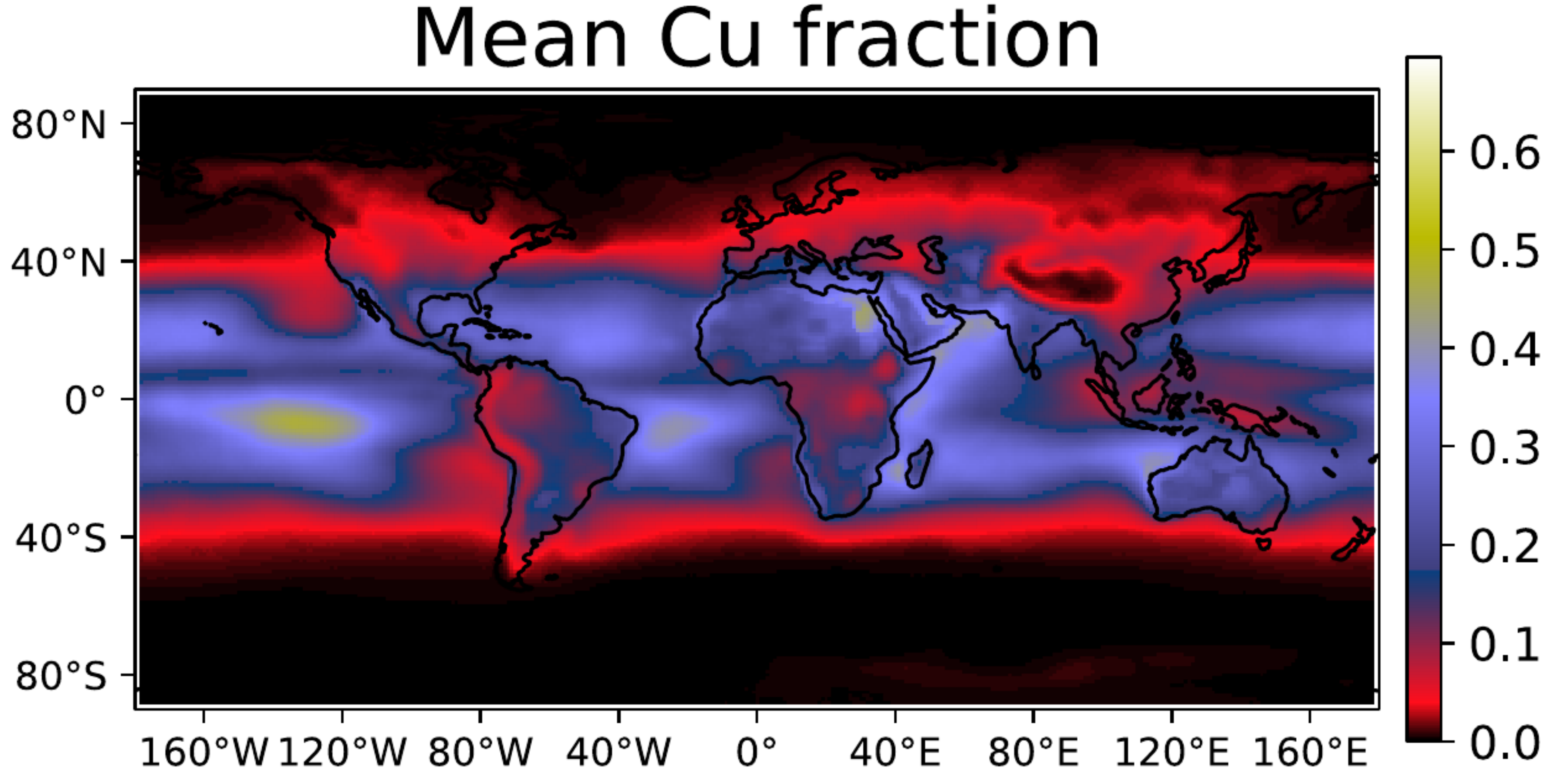}
        \subcaption{}
        \label{fig:Cu}
    \end{subfigure}\\
    \begin{subfigure}{0.49\linewidth}
    	\centering
        \includegraphics[width=\linewidth]{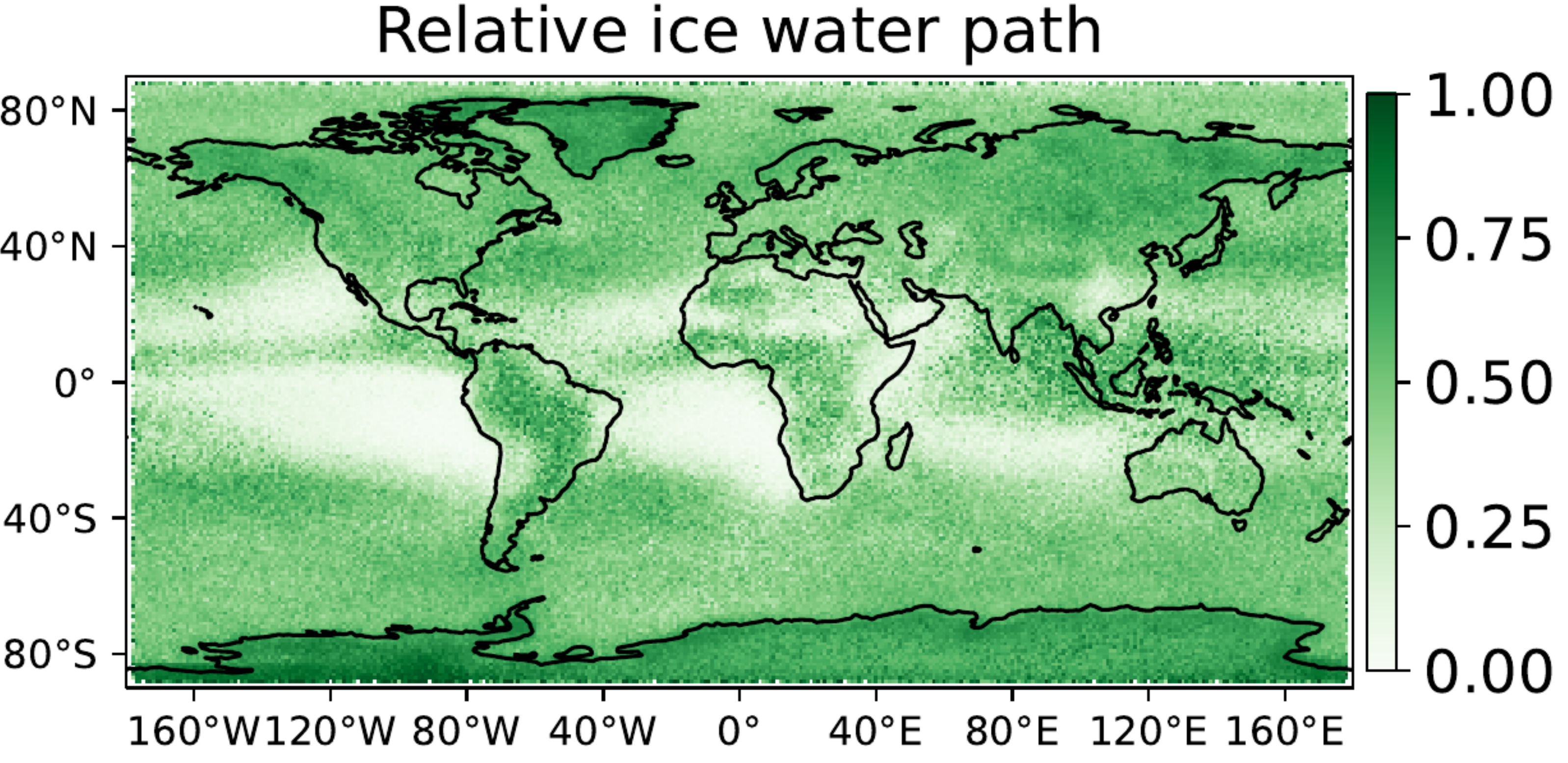}
        \subcaption{}
        \label{fig:ceri}
    \end{subfigure}
    \begin{subfigure}{0.49\linewidth}
    	\centering
        \includegraphics[width=\linewidth]{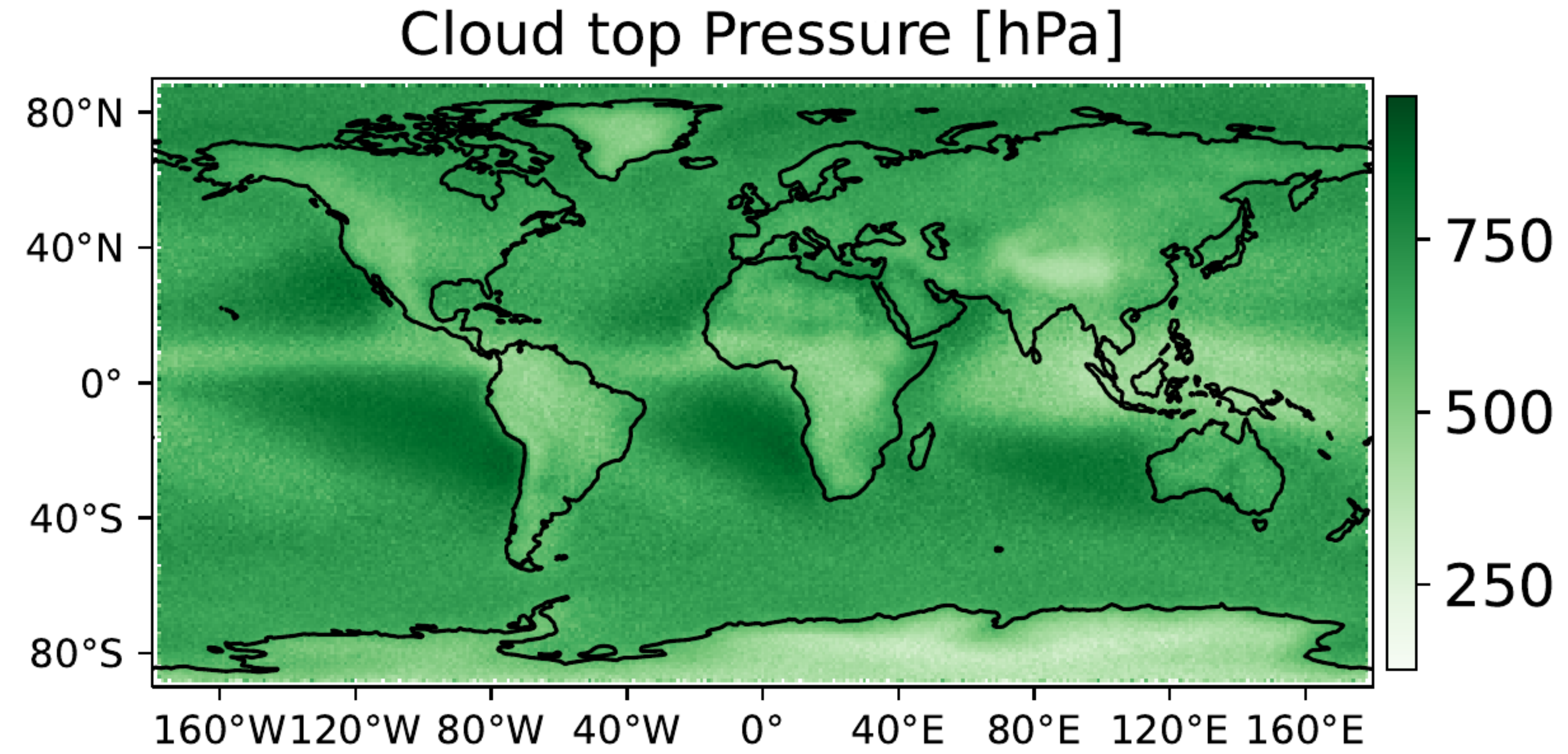}
        \subcaption{}
        \label{fig:ctp}
    \end{subfigure}
    \caption{Mean predicted class fractions for Ci(\ref{fig:Ci}) and Cu(\ref{fig:Cu}) compared to the mean feature values for \textit{ptop}(\ref{fig:ctp}) and the relative ice water path. Grid cells are 100$\times$100 pixels and the RF was trained on $(100\;\rm{km})^2$ grid cells using the optimal set of features.}
    \label{fig:compare_inputs}
\end{figure*}

We can perform similar analysis on the results obtained using the ESACCI data, also taking into account regional prevalence of certain features. Fig.~\ref{fig:compare_inputs} shows global maps of predictions for the Ci and Cu classes and the features \textit{ceri} and \textit{ptop}. We find a high Cu fraction particularly in the subtropical subsidence regions characterized by a high average cloud top pressure. These areas contain essentially no predicted Ci clouds. In contrast, Ci clouds are frequently predicted in low- and mid-latitude regions and are characterized by a large ice content, visible as the large mean values of \textit{ceri}$>15$\;µm. This is consistent with the expectation that cirrus clouds are characterized by high cloud tops and high relative ice content, whereas the opposite applies to shallow cumulus. Indeed, the fraction of Ci is slightly positively correlated with \textit{ceri} (Pearson correlation $c_P =0.18$) and anti-correlated with \textit{ptop} ($c_P=-0.56$), while the opposite is true for Cu (\textit{ceri}: $c_P=-0.4$, \textit{ptop}: $c_P=0.29$). This is another indicator for the physical consistency of the predicted class fractions with the WMO cloud types.

\section{Feature Importance \label{sec:feature}}
Beyond the physical relationship of the classes to the features, we can also determine which features are important for the model to provide good predictions. We did this for the regression model, because this will be the model that is applied on new datasets, where not all features might be available. Note that the following analysis applies only to this specific model. A model trained on fewer features might rely on a different combination of these.

\begin{figure*}
	\centering
	\begin{subfigure}{0.45\textwidth}
	    \includegraphics[width=\linewidth]{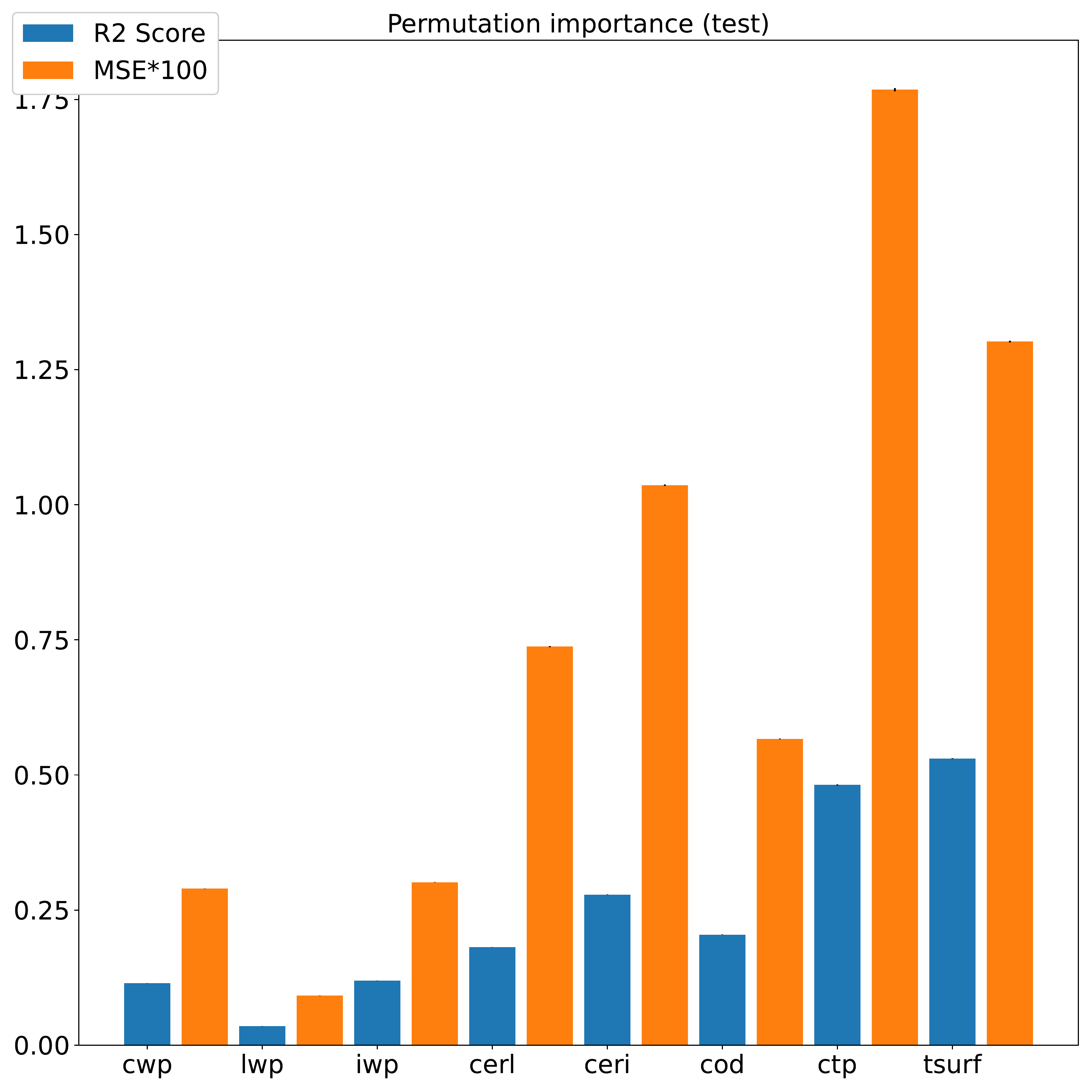}
	    \caption{\label{fig:testpmi}}
	\end{subfigure}
    \begin{subfigure}{0.45\textwidth}
	    \includegraphics[width=\linewidth]{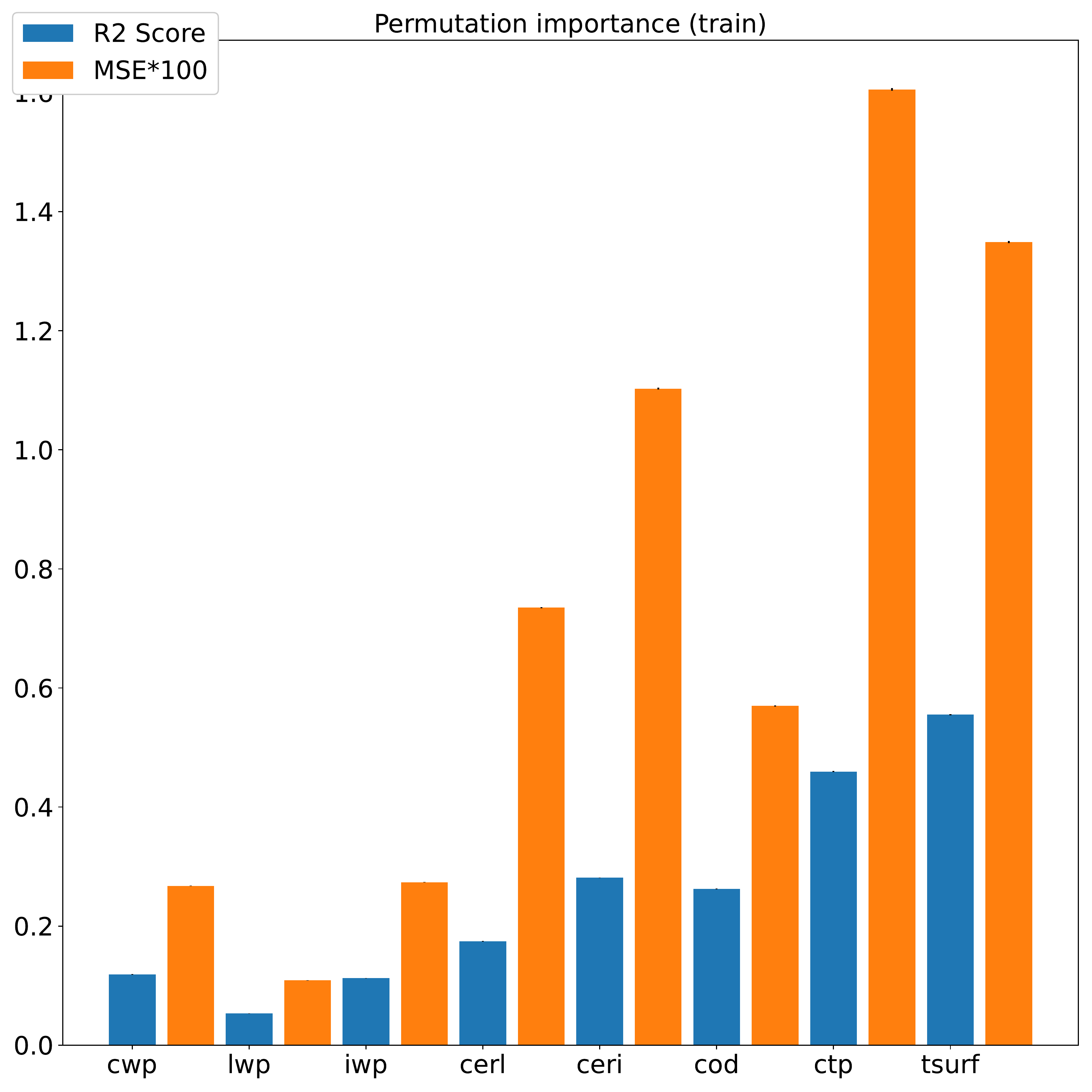}
	    \caption{\label{fig:trainpmi}}
	\end{subfigure}\\
    \begin{subfigure}{0.45\textwidth}
	    \includegraphics[width=\linewidth]{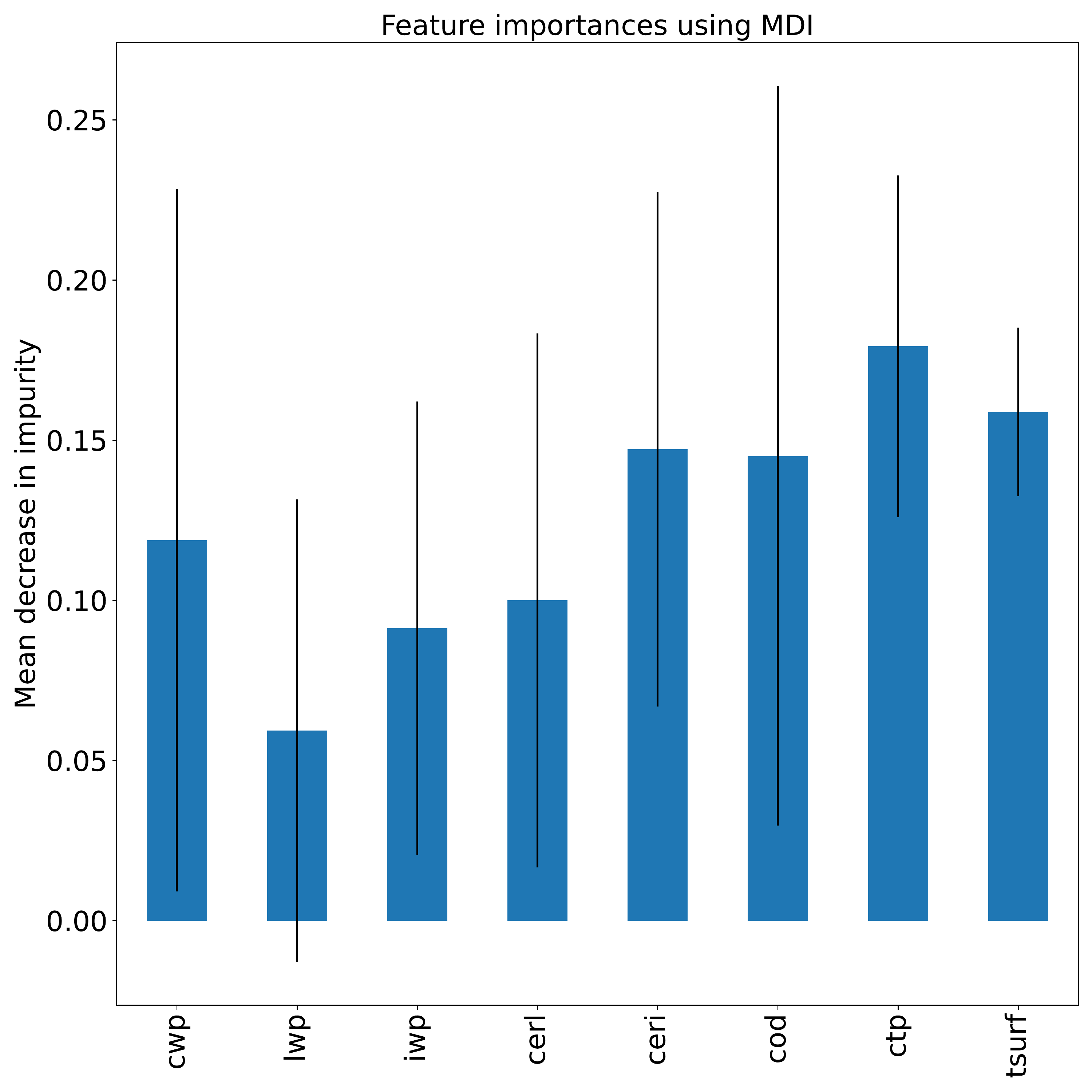}
	    \caption{\label{fig:mdi}}
	\end{subfigure}
    \begin{subfigure}{0.45\textwidth}
	    \includegraphics[width=\linewidth]{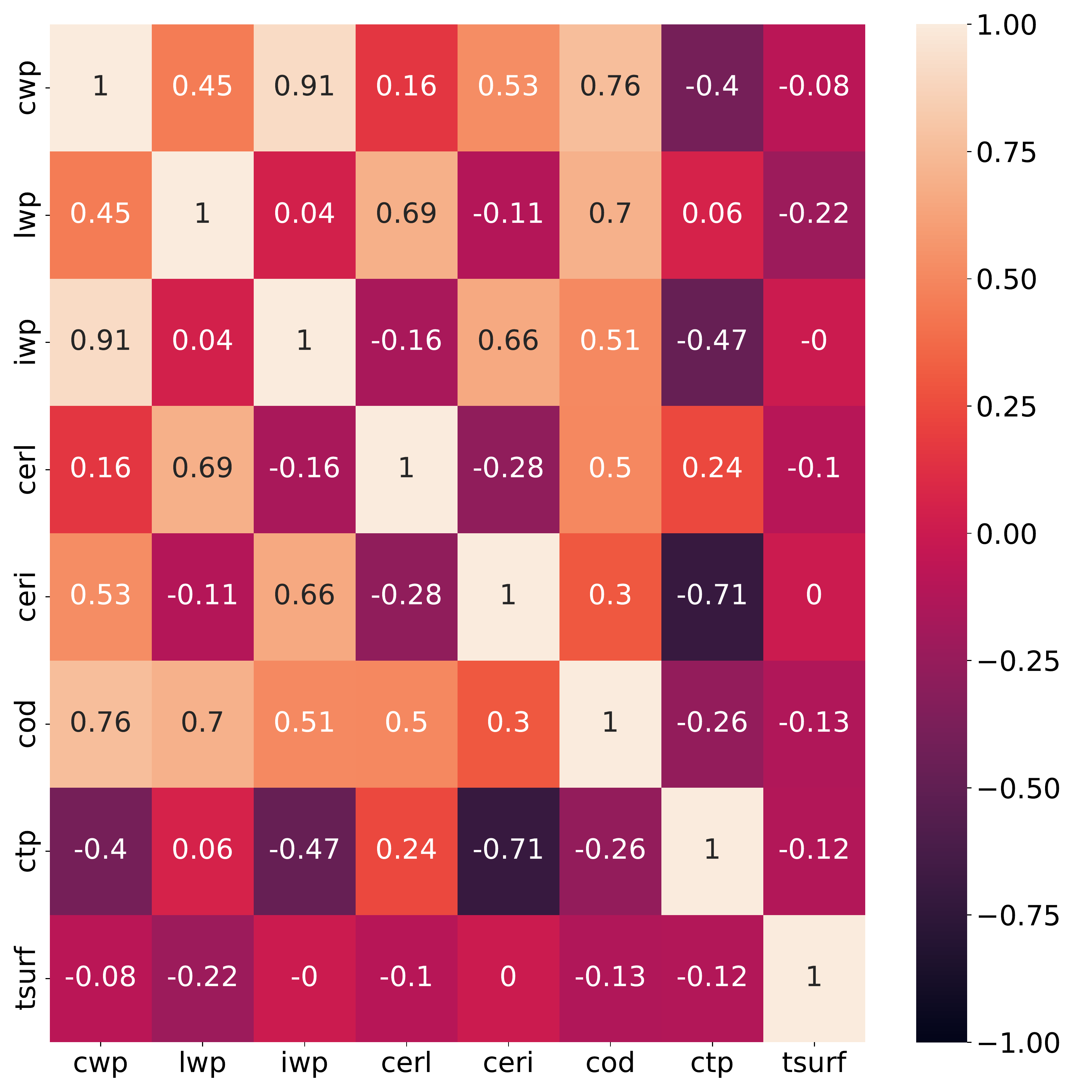}
	    \caption{\label{fig:correlation}}
	\end{subfigure}
    \caption{Analysing the impact the individual features used in the regression have on the predictions of the RF. \ref{fig:testpmi}: Permutation importance, in terms of the R2-Score and the mean squared error (scaled for visibility), computed on the test split of the data. Standard deviation is shown but so small it is barely visible. \ref{fig:trainpmi}: same as \ref{fig:testpmi} but computed on the train split. In both cases the importance was computed for a dataset of one million samples, using 20 different permutations per feature. \ref{fig:mdi}: Mean decrease in impurity with standard deviation over all trees. \ref{fig:correlation}: Pearson correlation of the features \label{fig:permimp}.}
\end{figure*}

First, we analysed the features of the regression model using the permutation importance method. This method quantifies the impact of shuffling individual features throughout the data while keeping all other features fixed. The importance is measured in terms of the decrease of chosen metrics, of which we use both the R2-score and the mean squared error (MSE). These metrics are displayed in Fig.~\ref{fig:permimp} for an RF trained on $(100\;\rm{km})^2$ grid cells. The permutation importance can both be computed on the train (Fig.~\ref{fig:trainpmi}) and test data (Fig.~\ref{fig:testpmi}). A high importance in the train split can indicate that the model overfits on the respective feature. The importance in the test split highlights the features important for generalization. However, for our case the permutation importance is virtually identical for both splits, indicating good generalization to the test set. The features \textit{tsurf} and \textit{ptop} seem to have the largest impact on both the R2-Score and the MSE when permuted, followed by \textit{ceri} and \textit{cod}.

The permutation importance can be skewed when features are correlated, as information about the permuted feature can be inferred from its correlated values. As Fig.~\ref{fig:correlation} shows, all features but \textit{tsurf} are strongly correlated with at least one other feature ($|c_P |\geq 0.69$). This might lead to a comparatively high feature importance for \textit{tsurf}. However using correlated features has still proven to be useful to the results. Even though most of the features are correlated with at least one other feature, these correlated features provide additional information that would otherwise be lost when spatially averaged.

We additionally took into account the mean decrease in impurity (MDI) attributable to the individual features (Fig.~\ref{fig:mdi}). The importances produced this way are similar to those indicated by the permutation importance using MSE. 

Taken together, these feature importance measures indicate that the model relies strongly on \textit{ptop} for its predictions. The results for the other features are less clear but a strong dependence on \textit{ceri} and \textit{cod} is likely, since their impact ranks high across all measures. Due to the correlated features and the fact that the MDI can only be determined for the test set these results however only give a rough indication about which features are required for successful application of our method to climate models. The high importance these measures attribute to \textit{tsurf} is however unexpected.

\section{Impact of changing the coarse graining resolution of ESACCI\label{sec:coarse}}
\begin{figure*}
    \centering
    \includegraphics[width=\linewidth]{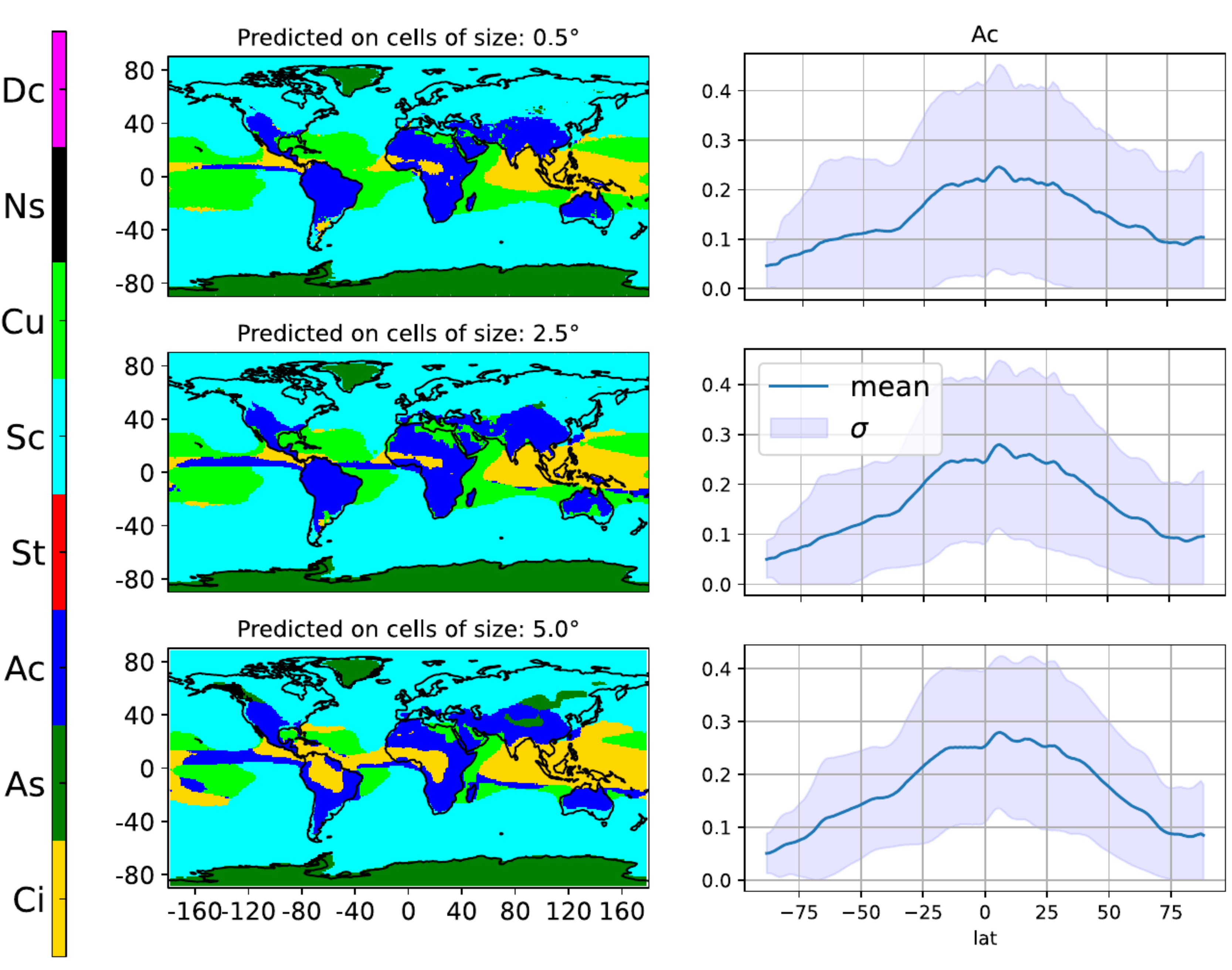}
    \caption{Results for predictions performed on ESACCI data with different coarse graining resolutions. The panels on the left show the most common cloud type in each $1^{\circ}\times 1^{\circ}$ area, while those on the right show the zonal average for the Ac class.}
    \label{fig:coarsecomp}
\end{figure*}

The MODIS and ESACCI data can not be coarse grained such that corresponding grid cells everywhere on the globe cover the same area. This could be achieved by interpolative regridding, which we want to avoid here. We instead aim to find a coarse grained resolution that is most similar to the other across the globe and assume that the grid cell averaging mitigates resulting differences. We have therefore evaluated the RF trained on a fixed grid cell size from coarse grained MODIS on different resolutions of the ESACCI data. We applied an RF trained on $(100\;\rm{km})^2$ grid cells on $(0.5^{\circ})^2, (2.5^{\circ})^2$ and $(5^{\circ})^2$ ESACCI grid cells.  Fig.~\ref{fig:coarsecomp} compares these results using the most common cloud type, binned to a $1^{\circ}\times 1^{\circ}$ grid as well as the zonal average of the As class. Note that again, the cloud distributions per coarse grained grid cell are interpreted as point values for the grid cell center. The representation using the most common cloud type, while being useful to represent the result in a single plot, hides the less prevalent cloud types, such that only Sc (teal), Ci (gold), As (dark green), Ac (blue) and Cu (light green) are displayed . The results show, that the global class distributions stay similar, but some features shift depending on the resolution. The other cloud types do however not disappear completely in the predictions but are being ``hidden'' in this representation. For example, we see that the Cu class gets hidden by Ci when the resolution is decreased. An interesting difference is the increase in As fraction in central Asia with decreasing resolution. Furthermore, the zonal averages of Ac show that with decreasing resolution the variability of the predictions decreases as well. The zonal mean of Ac however stays the same for all three resolutions.

We conclude that if trained on data coarse grained to around $(100\;\rm{km})^2$, the method can provide useful predictions on any resolution typical for ESMs. Even though, some classes with similar regional occurrence might ``switch places'', the zonal distributions stay essentially the same across resolutions indicating no fundamental changes.

\section{Impact of temporal resolution\label{sec:MM}}  
\begin{figure*}
	\centering
	\begin{subfigure}{0.49\linewidth}
    	\centering
        \includegraphics[width=\linewidth]{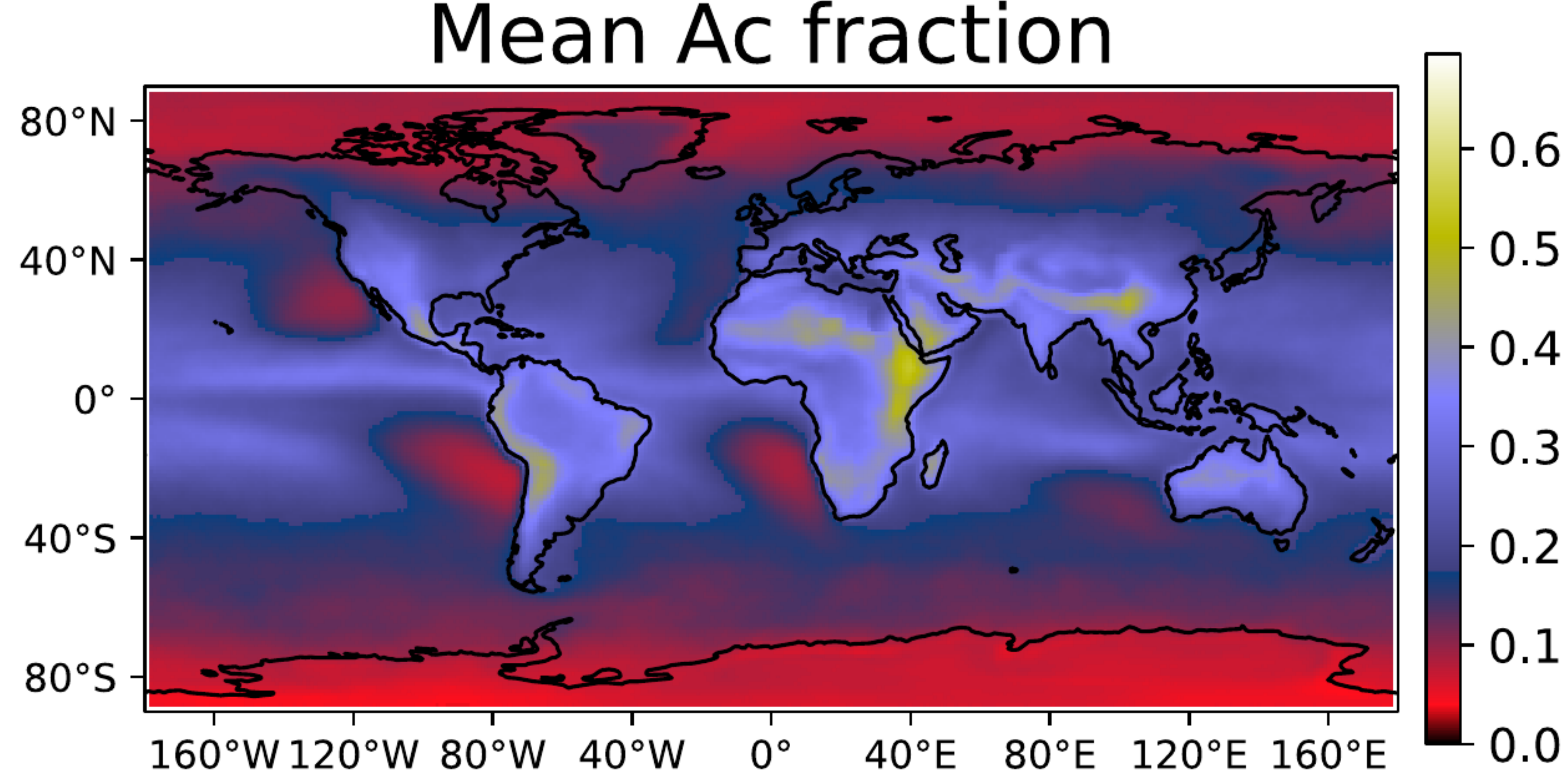}
        \subcaption{}
        \label{fig:Ac_de}
    \end{subfigure}
    \begin{subfigure}{0.49\linewidth}
    	\centering
        \includegraphics[width=\linewidth]{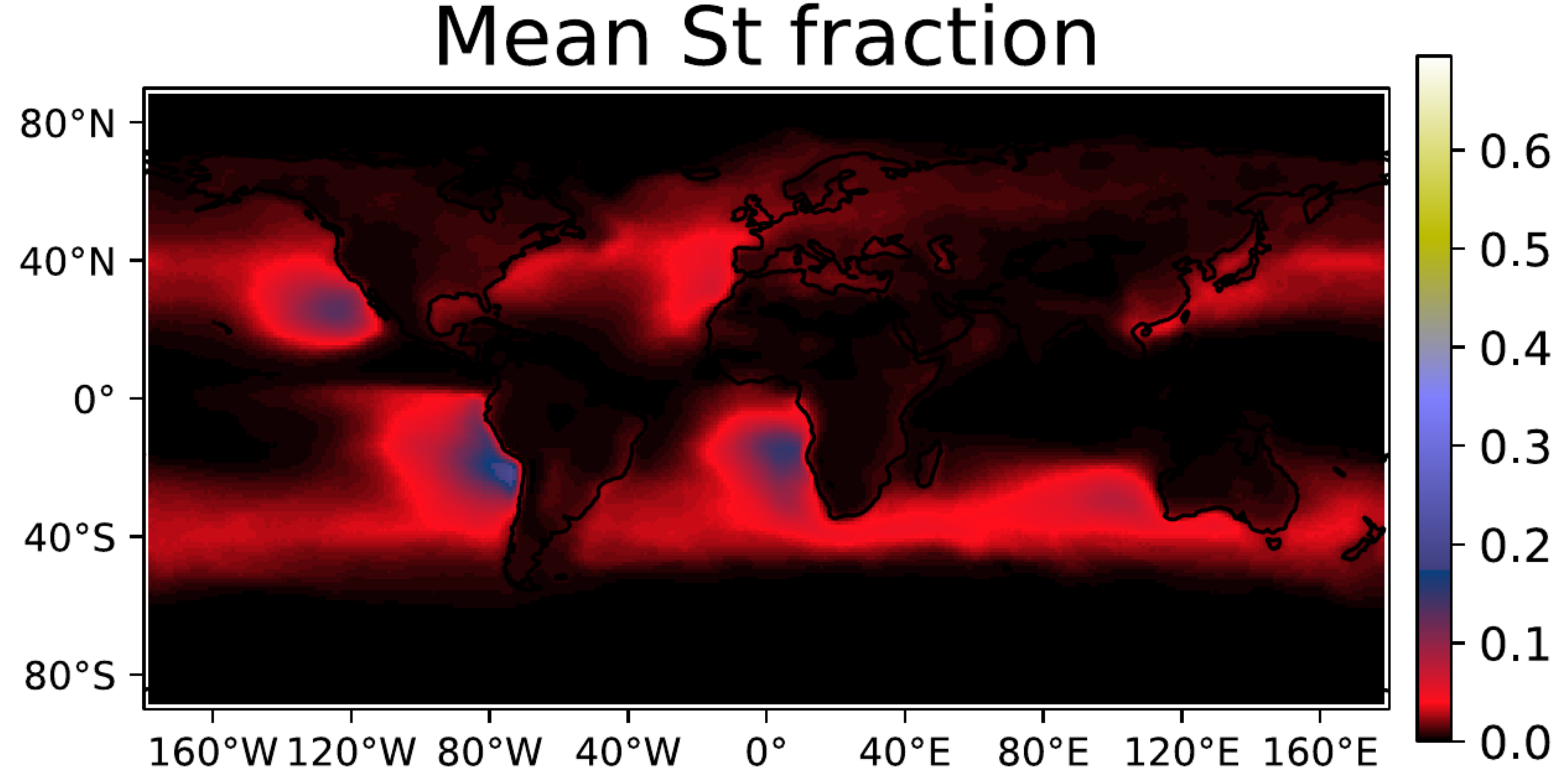}
        \subcaption{}
        \label{fig:St_de}
    \end{subfigure}\\
    \begin{subfigure}{0.49\linewidth}
    	\centering
        \includegraphics[width=\linewidth]{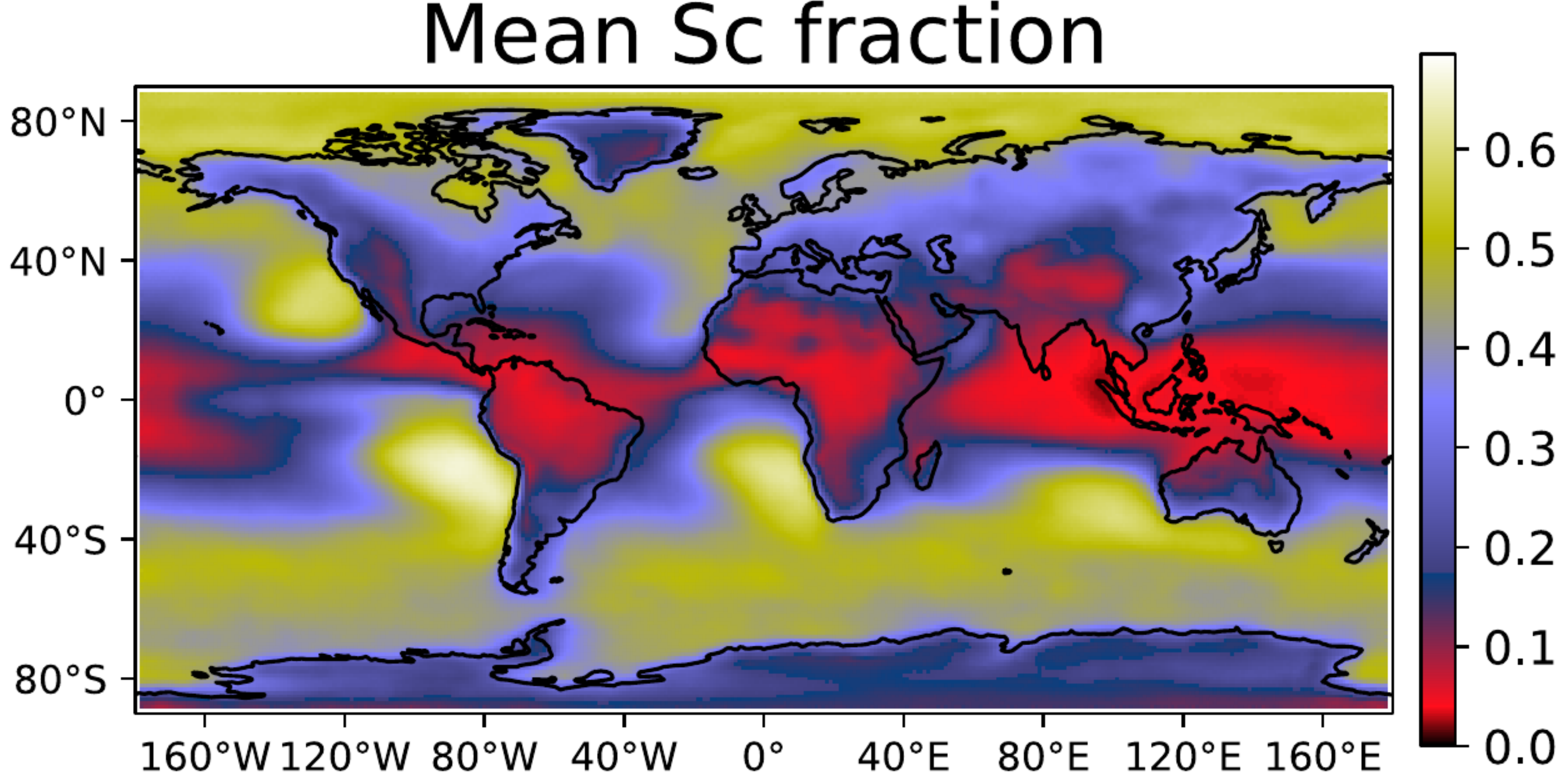}
        \subcaption{}
        \label{fig:Sc_de}
    \end{subfigure}
    \begin{subfigure}{0.49\linewidth}
    	\centering
        \includegraphics[width=\linewidth]{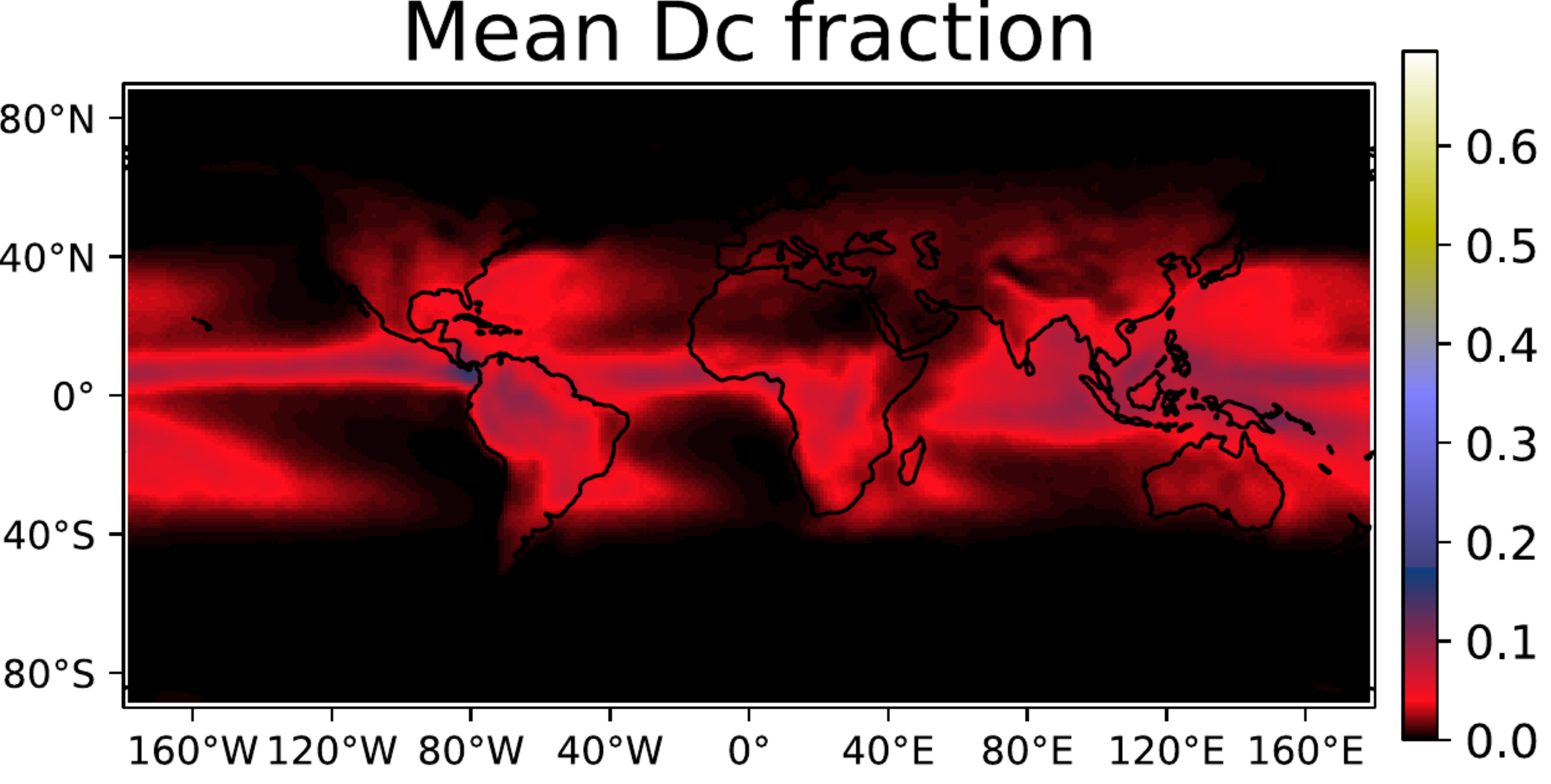}
        \subcaption{}
        \label{fig:Dc_de}
    \end{subfigure}
    \caption{Examples of predicted mean class fractions using feature values from instantaneous source data. Predictions were made using an RF trained on $(100\;\rm{km})^2$ grid cells applied on $100\times 100$ ESACCI pixels.
    \label{fig:allc_inst}}
\end{figure*}

\begin{figure*}
	\centering
    \begin{subfigure}{0.49\linewidth}
    	\centering
        \includegraphics[width=\linewidth]{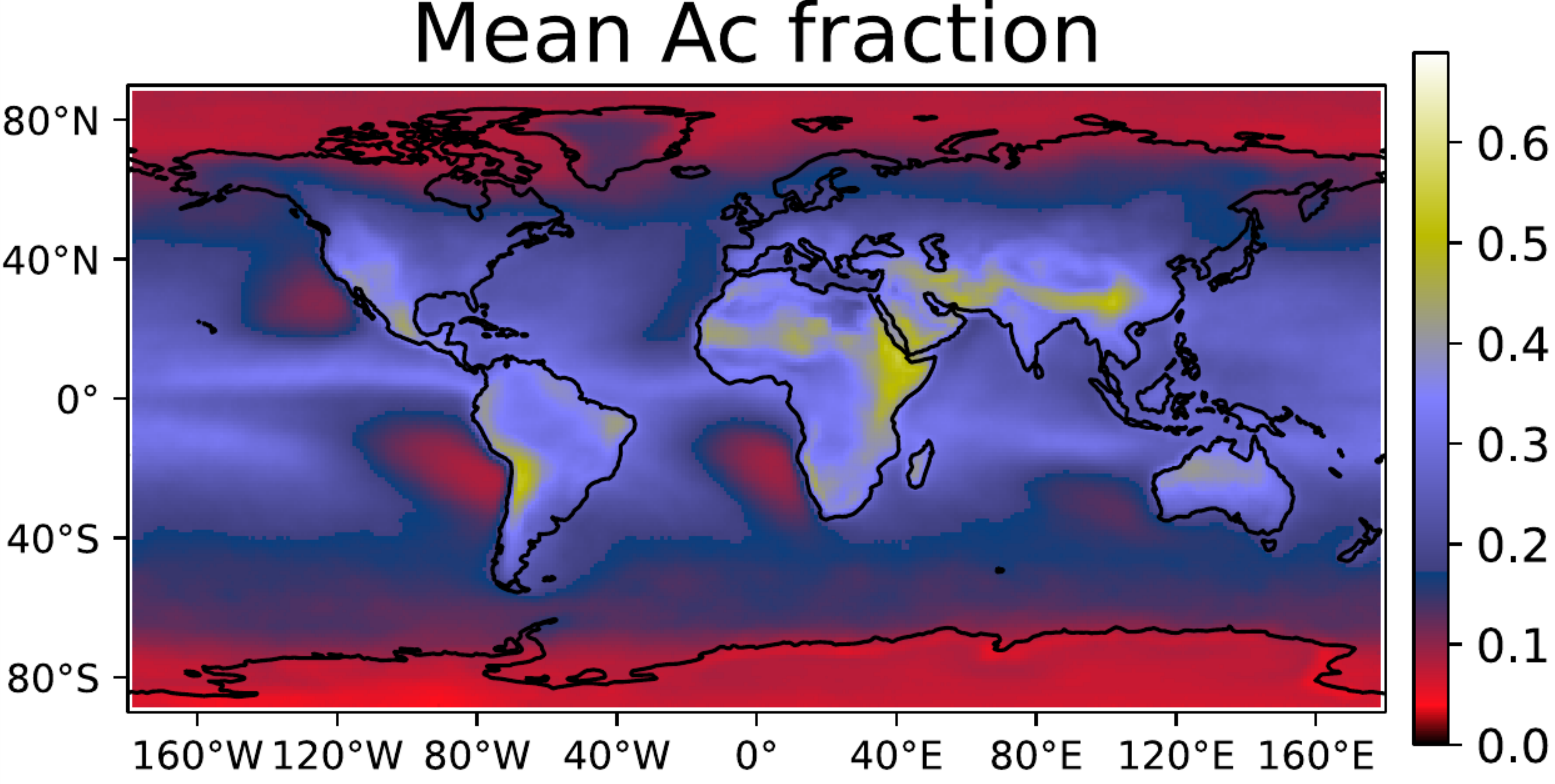}
        \subcaption{}
        \label{fig:Ac_dn}
    \end{subfigure}	
    \begin{subfigure}{0.49\linewidth}
    	\centering
        \includegraphics[width=\linewidth]{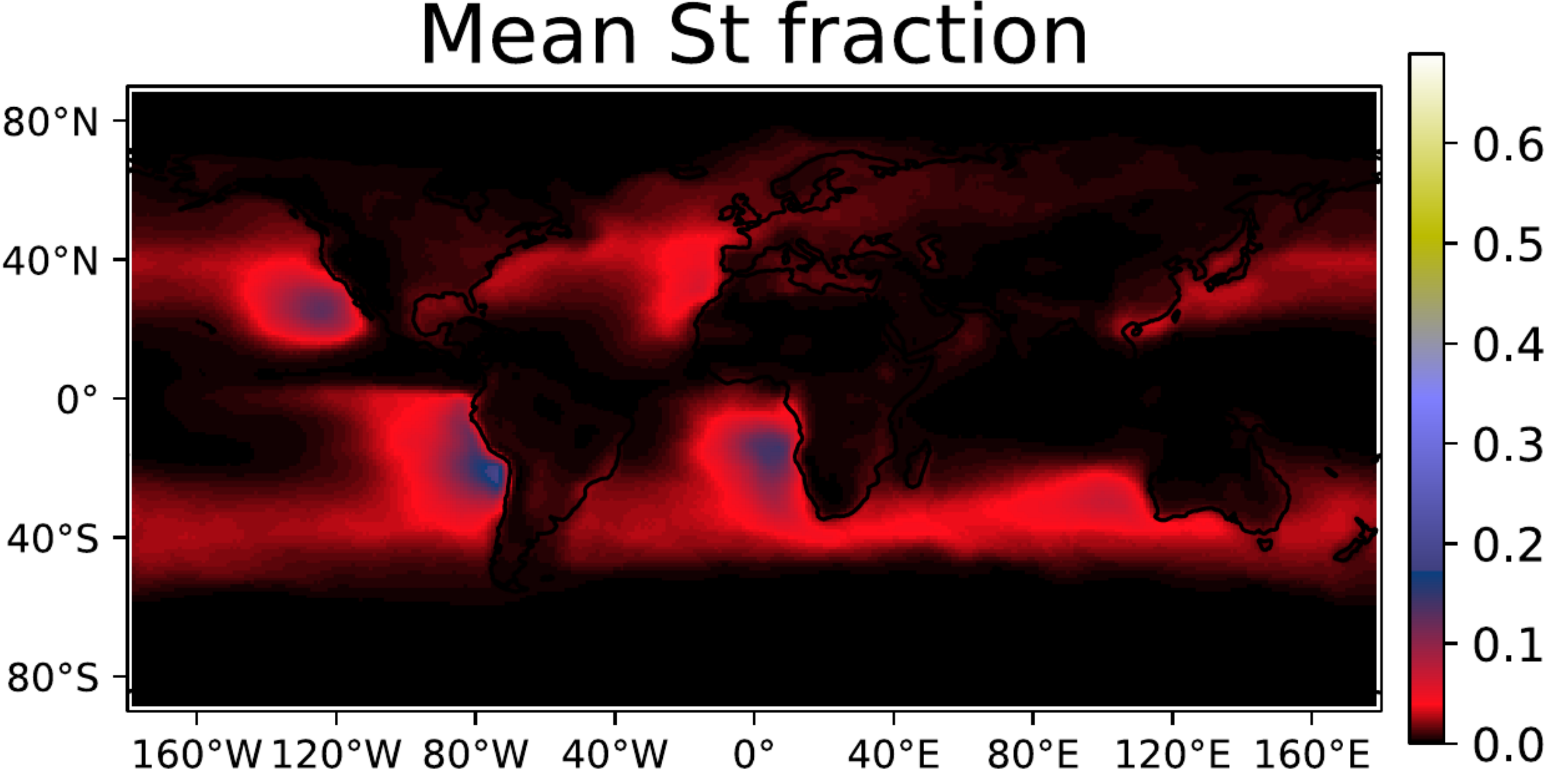}
        \subcaption{}
        \label{fig:St_dn}
    \end{subfigure}\\
    \begin{subfigure}{0.49\linewidth}
    	\centering
        \includegraphics[width=\linewidth]{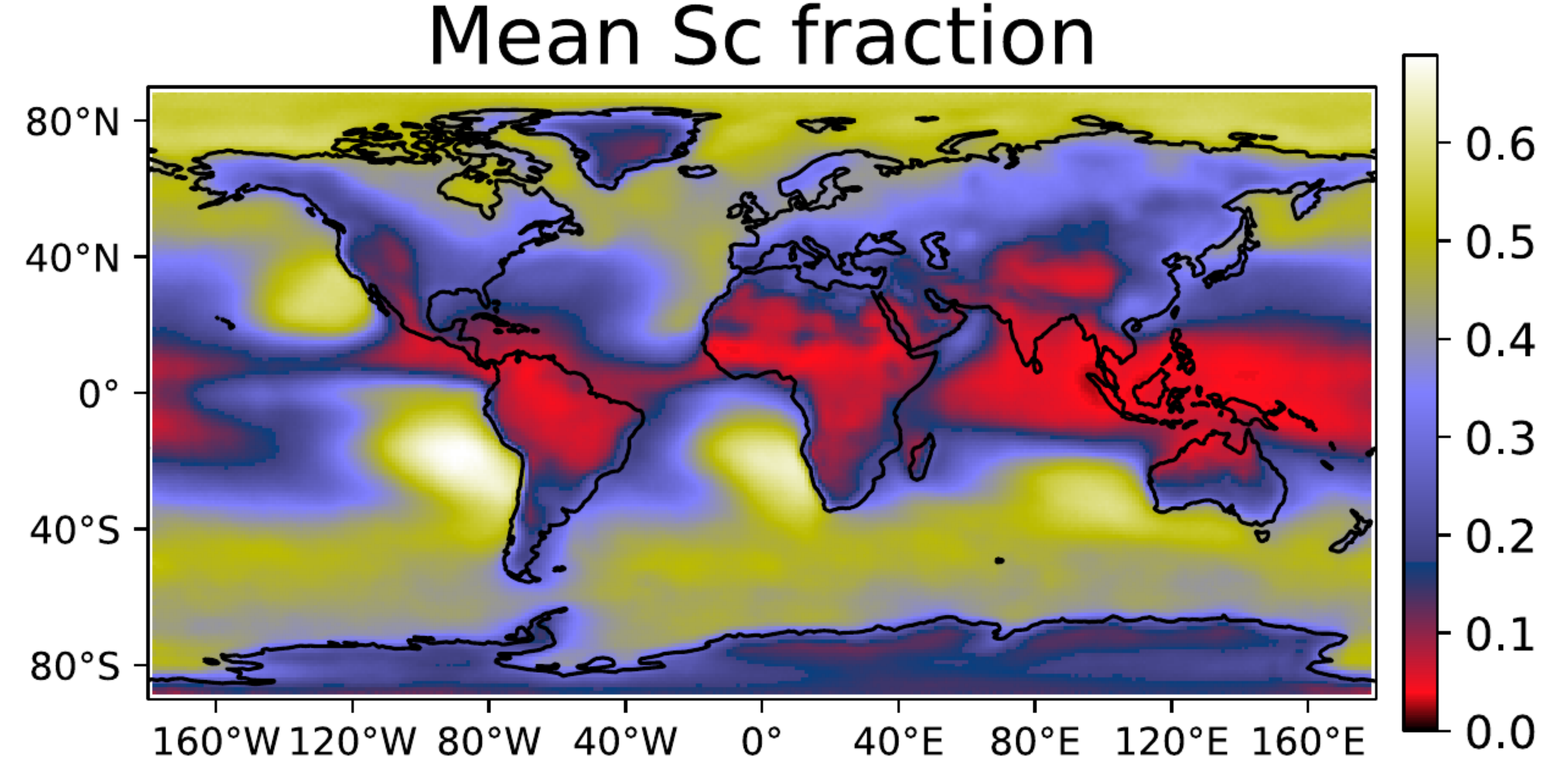}
        \subcaption{}
        \label{fig:Sc_dn}
    \end{subfigure}
    \begin{subfigure}{0.49\linewidth}
    	\centering
        \includegraphics[width=\linewidth]{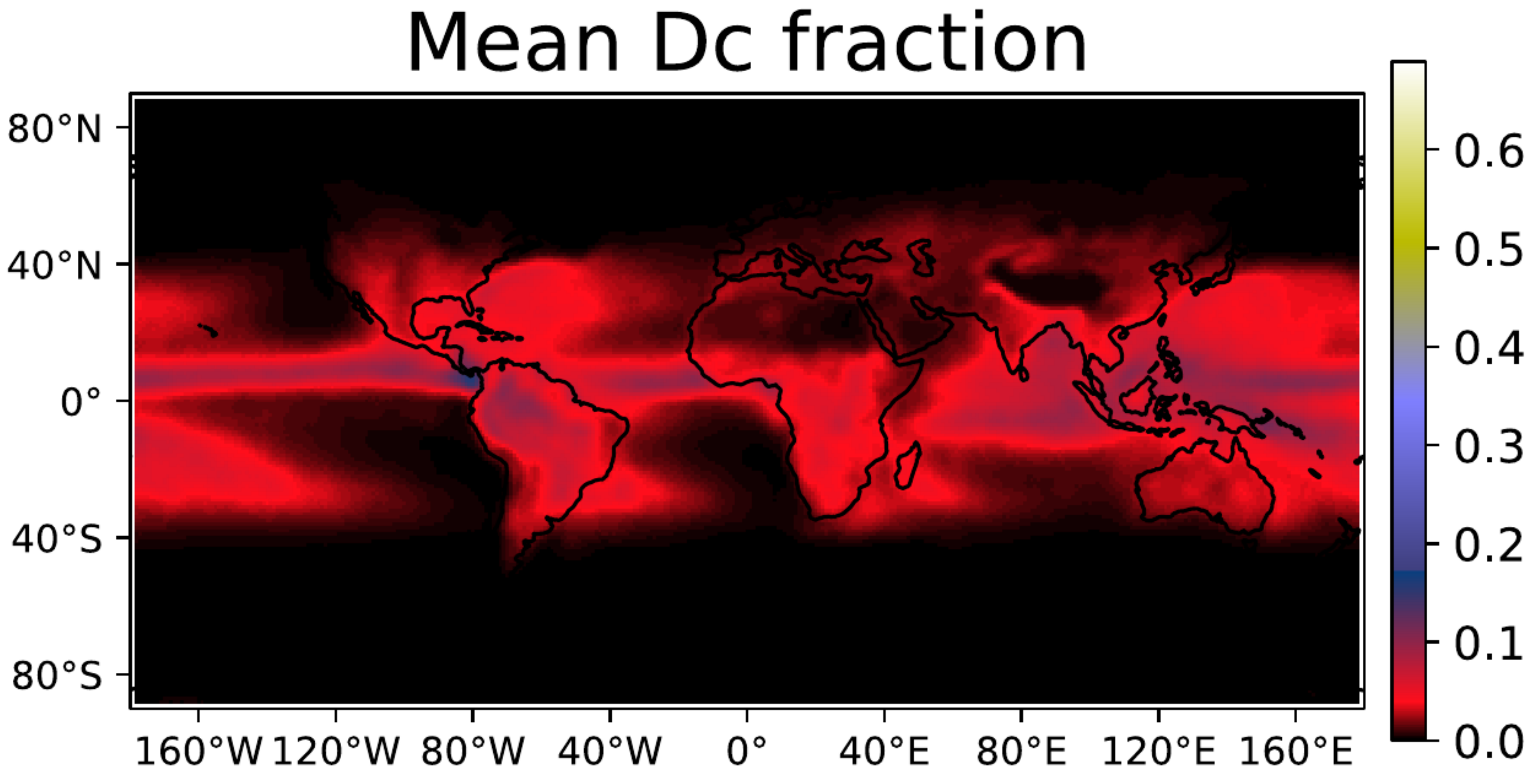}
        \subcaption{}
        \label{fig:Dc_dn}
    \end{subfigure}
    \caption{Mean class fractions using feature values obtained by averaged over ascending and descending orbits. Predictions were made using the same RF as for Fig.~\ref{fig:allc_inst}, using $100\times 100$ ESACCI pixels as well \label{fig:allc_mean}.}
\end{figure*}

The results shown in the main paper were produced by training and applying the machine-learning models to instantaneous data. Climate models, however, often provide output in the form of daily or monthly averages. This averaging process might introduce deviations from the feature distributions obtained with instantaneous data. The impact of using temporally averaged data is investigated in the following using averaged inputs for the regression model. Here, we use the two daily measurements from the ascending and descending orbits provided by the ESACCI dataset. We compare the results to the ones obtained with the same regression model applied to instantaneous data. Exemplary results for using features from the instantaneous and mean input data are shown in Fig.~\ref{fig:allc_inst} and Fig.~\ref{fig:allc_mean}, respectively. For all classes, the distributions look very similar, also to the results obtained using an RF trained and applied on higher resolution data (Fig.~\ref{fig:allctypes_higres} in the main paper). The per class correlations range between $c_P^{Dc}=0.714$ and $c_P^{As}=0.893$, with unweighted mean $\overline{c_P}=0.839$, this giving an indication that using daily averages instead of instantaneous features for this method is possible. 

\begin{figure*}
    \centering
    \includegraphics[width=\linewidth]{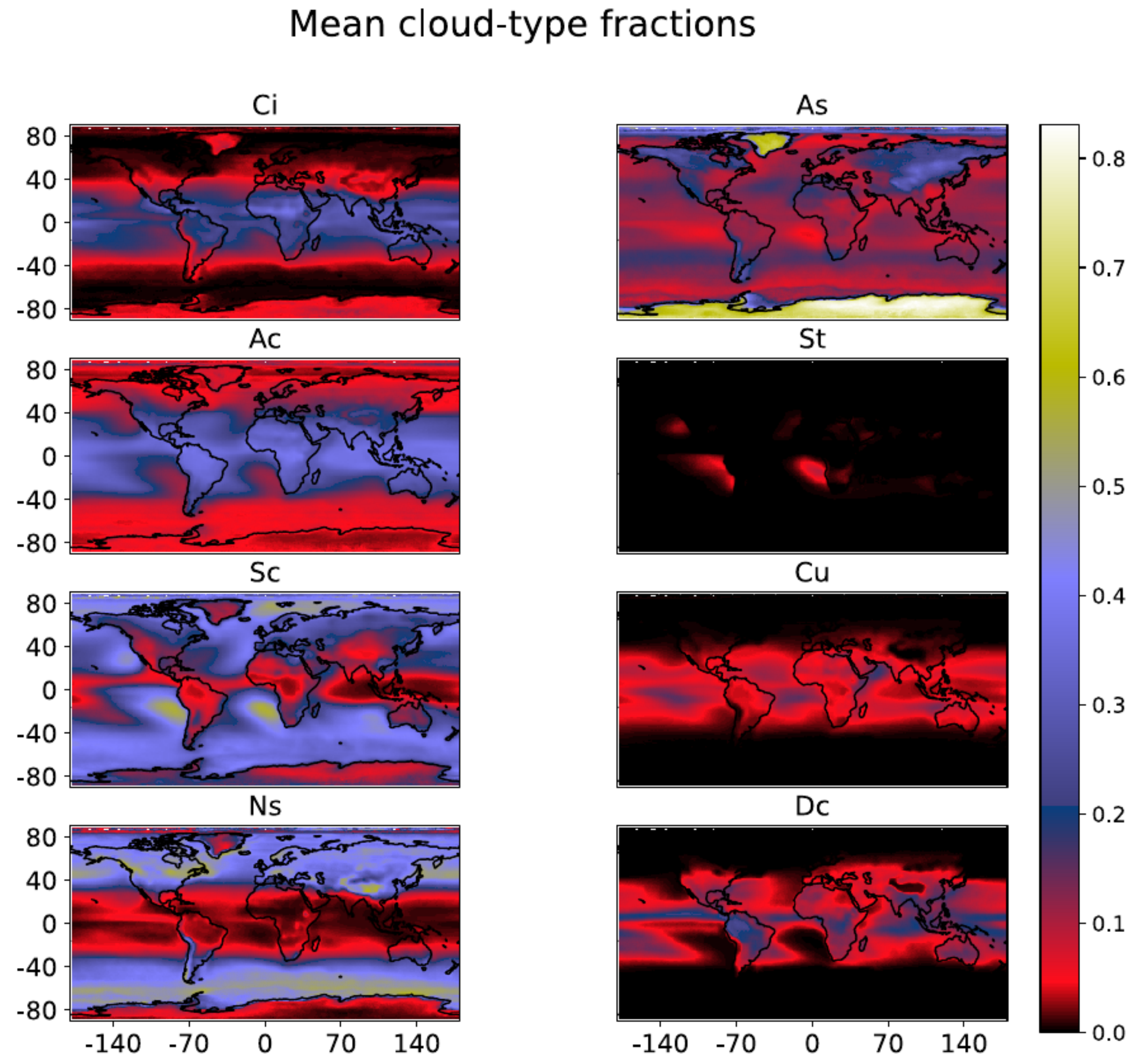}
    \caption{Average predicted fractions using features obtained from monthly mean data (L3C), using the RF trained on $(100\;\rm{km})^2$ grid cells. The source data consists of 220 months sampled randomly between 1984 and 2016. The grid cells are constructed from 10$\times$10 ESACCI pixels of $0.5^{\circ}$ resolution.}
    \label{fig:mn_all}
\end{figure*}

We applied the RF model trained on $(100\;\rm{km})^2$ grid cells to ESACCI monthly mean (L3C) data. Fig.~\ref{fig:mn_all} shows that the average predicted fractions are similar to those of the instantaneous input data (Figures \ref{fig:allc_inst}, \ref{fig:allc_mean}), but with less pronounced geographical features. As one would expect from using monthly means, the predicted fractions appear smoothed out and show rather similar magnitudes over large areas. The most notable difference in the representation of the individual classes with respect to using instantaneous data is the globally increased amount of the predicted Ns fraction. We see a further increase in Sc and a decrease in St compared to the distribution in CloudSat, suggesting that monthly mean data are not well suited as an input to our method. The increase in both Ns and As further suggests that the monthly average data show much larger effective ice particle radii than the instantaneous data. In fact we find a 4 fold increase in the median ice particle radius and similarly large increases in the cloud water path. Since both, the L3C and L3U ESACCI data, are derived from the same instrument, the time averaging of the data must somehow cause this increase, which in turn causes the regression to produce unreliable results.

